\pdfoutput=1 
\documentclass[11pt]{article}
\usepackage[utf8]{inputenc}
\usepackage[DIV12]{typearea}
\usepackage{amsmath,amsfonts,amssymb,amscd}
\usepackage{graphicx}
\usepackage{cite}
\usepackage{mathrsfs}
\usepackage{bbm}
\usepackage{soul,xcolor}
\usepackage{ulem}
\usepackage[bookmarks=false]{hyperref}
\hypersetup{
    pdftitle = {Extra Dimensions Beyond the Horizon},
    pdfauthor = {Jeffrey Kuntz, Andreas Trautner},
    colorlinks=true,
    linkcolor=blue,
    filecolor=blue,      
    urlcolor=blue,
    citecolor=blue
}

\usepackage{physics}
\usepackage{tensor}
\usepackage{indentfirst}
\usepackage[font=footnotesize,labelfont=bf]{caption}
\usepackage{float}
\usepackage{cancel}

\newcommand{\Lag}{\ensuremath{\mathcal{L}}}
\newcommand{\Ord}{\ensuremath{\mathcal{O}}}

\newcommand{\Reals}{\ensuremath{\mathbb{R}}}

\newcommand{\nreturn}{\nonumber \\[0.4em]}
\newcommand{\return}{\\[0.8em]}
\newcommand{\Nf}{\ensuremath{\mathcal{N}}}
\newcommand{\Af}{\ensuremath{\mathcal{A}}}
\newcommand{\Bf}{\ensuremath{\mathcal{B}}}
\newcommand{\hKK}[1]{\hat{h}_{{}_{\scriptstyle #1}}{}}
\newcommand{\hKKp}[1]{\hat{h}'_{{}_{\scriptstyle #1}}{}}

\newcommand{\e}{\ensuremath{\mathrm{e}}}

\DeclareMathOperator\arcsinh{arcsinh}

\definecolor{darkgreen}{HTML}{109930}

\setstcolor{red}

\def\mytitle{
Extra Dimensions Beyond the Horizon}
\title{\mytitle}

\begin{document}

\begin{titlepage}

\vspace*{2.0cm}

\begin{center}
{\LARGE\textbf{\mytitle}}

\renewcommand*{\thefootnote}{\fnsymbol{footnote}}

\vspace{1.2cm}
\large
Jeffrey Kuntz\footnote[1]{
\href{mailto:jkuntz@mpi-hd.mpg.de}{jkuntz@mpi-hd.mpg.de}
} 
and
Andreas Trautner\footnote[2]{
\href{mailto:trautner@mpi-hd.mpg.de}{trautner@mpi-hd.mpg.de}
}
\normalsize
\\[5mm]
\textit{
Max-Planck-Institut f\"ur Kernphysik \\ Saupfercheckweg 1, 69117 Heidelberg, Germany
} 
\end{center}
\vspace*{6mm}
\begin{abstract}
We discuss an extra-dimensional braneworld with a 5th dimension compactified on a circle.
As a characteristic feature, the warp factor is hyperbolic and separates the hidden and visible branes by a bulk horizon without a singularity. The two most widely separated scales of 4D physics -- the 4D Planck mass and 4D cosmological constant -- are determined by two physical scales in the extra dimension, namely: $(i)$ the proper size of the extra dimension, $R$, and, $(ii)$ the distance between the visible brane and the horizon, $R_0$. A realistic scale hierarchy between 4D Planck mass and 4D cosmological constant is obtained for $R/R_0\sim2.34$. The usual fine tuning is not reduced but promoted to a fine tuning of two separate brane energy densities that must approach the fundamental scale of the model with very high precision. Our scenario is based on an exact solution to the 5D Einstein equations with a strictly empty bulk and Friedmann-Lema\^itre-Robertson-Walker metric on the 4D branes. This requires positive 4D brane energy densities and describes an adiabatic runaway solution in agreement with the de Sitter swampland conjecture.
The Kaluza-Klein~(KK) graviton states are solutions of a modified P\"oschl-Teller potential which permits a discrete graviton spectrum of \textit{exactly~two}~modes. In addition to the usual massless graviton, our scenario predicts an extra massive spin--2 graviton with a mass gap of $m_1=\sqrt{2}H_0\approx2\times10^{-33}\,\mathrm{eV}$ which might be detectable in the foreseeable future. A KK tower of gravitons, or a possible continuum of massive graviton states, is prohibited by unitarity with respect to the horizon. 
We discuss hurdles in turning this model into a realistic cosmology at all times, which points us towards 4D brane tensions that that must be raising towards the fundamental scale of the model, while the observable 4D expansion rate is decreasing.
\end{abstract}
\thispagestyle{empty}
\clearpage

\end{titlepage}

\section{Introduction}
Studying the phenomenology of models with extra spatial dimensions has a long history, see e.g.~\cite{Rubakov:1983bb}.
Well-known incarnations include the un-warped Large Extra Dimensions~\cite{Antoniadis:1998ig,Arkani-Hamed:1998jmv}, 
the power-law warped Linear Dilaton model~\cite{Antoniadis:2001sw,Antoniadis:2011qw,Cox:2012ee,Baryakhtar:2012wj}, 
and the exponentially warped Randall-Sundrum (RS) models~\cite{Randall:1999ee,Randall:1999vf}.
Extra-dimensional scenarios with orbifold geometry are motivated from string theory~\cite{Horava:1995qa,Lukas:1998yy}
and are a widely accepted way to conceptually address the electroweak scale hierarchy problem of the effective 4D theory. 

\begin{figure}[!t!]
\centering
\includegraphics[width=1.0\linewidth]{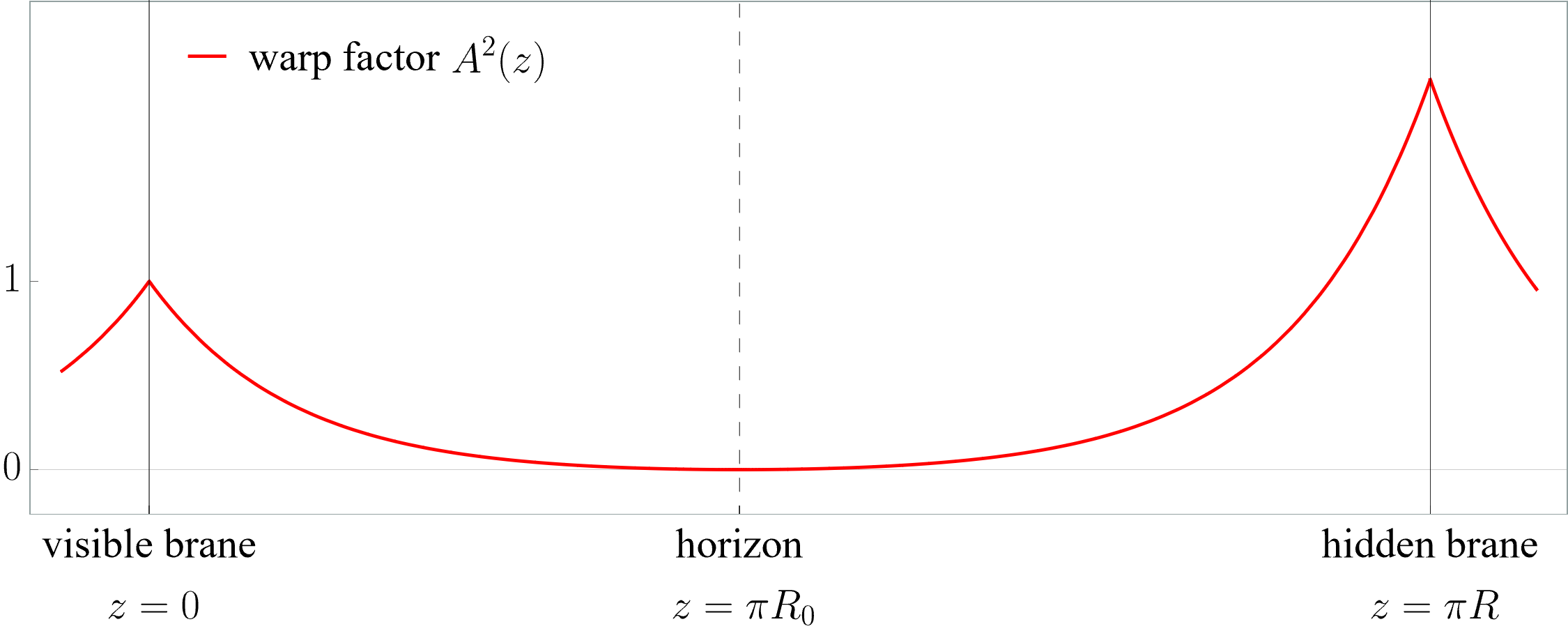}
\caption{\label{fig:Aplot}
Qualitative behavior of the warp factor (also corresponding to the shape of the metric background and zero mode) as a function of the extra dimensional coordinate $z$, here illustrated for $\mu=H_0$. The warp factor and first derivative are zero at $z=\pi R_0=c_z/\mu$ with the first derivative changing sign.
If $R>2R_0$, the barycenter of the curve lies behind the horizon, thereby explaining a suppressed effective 4D Planck mass on the visible brane.
Independently, the distance from the visible brane to the horizon, $\pi R_0$, determines the four-dimensional expansion rate $H_0$. Taking $\mu\sim\mathrm{TeV}$,
realistic values for 4D cosmological constant and Planck mass are obtained for $\pi\mu R_0\approx 104$ and $\pi\mu R\approx 243$, corresponding to $\gamma=R/R_0\approx 2.34$.}
\end{figure}

Very much like the original RS geometry, we consider a single compact extra dimension with orbifold geometry $S_1/\mathbbm{Z}_2$. 
This extra dimension is bounded by two three-dimensional branes located at $z=0,\pi R$. Without loss of generality we assume that the ``visible'' brane at $z=0$ houses the 4D particle physics Standard Model. In contrast to most models, we require a strictly empty bulk, i.e.\ require an exact solution of the $(55)$ component of the Einstein equations.
If both (brane and bulk) energy densities are assumed to be constant (which we will see to be an excellent approximation to more realistic cases that we will also discuss), the resulting metric has been previously discussed by Nihei~\cite{Nihei:1999mt}, Kaloper~\cite{Kaloper:1999sm}, and Kim and Kim~\cite{Kim:1999ja} (NKKK) and is given in our notation by
\begin{align} \label{NKKKmetIntro}
\mathrm{d}s^2 = g_{ab}\,\dd x^a \dd x^b = A^2(z)\Big(\dd t^2 - a^2(t)\delta_{ij}\dd x^i \dd x^j\Big) - \dd z^2 \,,
\end{align}
with the warp factor
\begin{align} \label{AwarpIntro}
&A^2(z) = \frac{H_0^2}{\mu^2}\sinh^2(\mu z - c_z)\;,\qquad\text{where}\qquad c_z := \mathrm{arcsinh}\left(\frac{\mu}{H_0}\right)\;,
\end{align}
as illustrated in in Fig.~\ref{fig:Aplot}. Here, $\mu^2:=-\Lambda/6M^3$ is a characteristic energy scale composed of the 5D anti-de Sitter (AdS) cosmological  
constant $\Lambda<0$ and 5D Planck mass $M$, while $H_0$ is the Hubble rate on the visible brane.

Instead of the exponential warping in the RS case, the NKKK metric has a \textit{hyperbolic} warp factor that touches zero at a location $z=\pi R_0$ in the bulk.
We will see that this zero corresponds to a causal horizon~\cite{Kaloper:1999sm}. In general, one may wonder whether any extra dimension with a horizon should be regarded as ending at that point~\cite{Kaloper:1999sm,Langlois:2000ns} (see also e.g.~\cite{Arkani-Hamed:1998cxo,Gomez:2000bu}).
However, in contrast to naively similar looking solutions such as the case of ``AdS-Schwarzschild'' \cite{Witten:1998zw,Horowitz:1998ha,Birmingham:1998nr,Chamblin:1999by,Kraus:1999it,Gubser:1999vj}, our solution does not exhibit a bulk curvature singularity and we will thus allow our spacetime to extend beyond the bulk horizon as was firstly suggested in \cite{Kaloper:1999sm} (to the best of our knowledge). We will show that continuing spacetime in this way is justified in the very same way as it is justified to extend 4D spacetime e.g.\ beyond a black hole horizon in the standard Schwarzschild solution, or beyond any causal horizon for that matter. Similar to how the gravitational effect of a black hole can be experienced despite the presence of an event horizon, we will demonstrate that also the gravitational influence of the hidden brane has measurable consequences on the visible brane. Namely, the graviton zero mode is warped such that the effective strength of gravity on the visible brane is substantially reduced. We will see that this is not in conflict with causality or unitarity. From a colloquial viewpoint we find it amusing that the formation and presence of bulk horizons could naturally explain why we cannot ``see'' extra dimensions.

\pagebreak
There are two relevant proper distance scales in the 5th dimension: $(i)$ the distance between visible brane and horizon, $R_0$, and $(ii)$ the distance between the visible and hidden brane corresponding to the overall size of the extra dimension, $R$. In the 4D effective theory, these two scales play an important role in the interpretation of the cosmological constant and effective Planck mass. The 4D Planck mass and the 4D Hubble expansion rate (in the cosmological constant dominated case) are found to be
\begin{align}
&M_p~\approx~ \sqrt{\frac{M^{3}}{\mu}}\,\mathrm{e}^{\,\mu\,\pi(R-2R_0)}~\approx~10^{18}\,\mathrm{GeV}& &\text{and}&
&H_0~\approx~2\,\mu\,\mathrm{e}^{-\mu\,\pi\,R_0}~\approx~10^{-33}\,\mathrm{eV}\;.&
\end{align}
Taking as a benchmark all fundamental scales of the order $M\sim|\Lambda|^{1/5}\sim\mu\sim v_{\mathrm{EW}}\sim\mathrm{TeV}$, the proper lengths of $R_0$ and $R$ in 5D are comparable and given by $\mu\pi R_0\approx104$ and $\mu\pi R\approx243$. 

The strict emptiness of the bulk prohibits any form of bulk potentials necessary for moduli stabilization. Our solution, therefore, must corresponds to a 5D runaway solution.
We will demonstrate that this runaway can be \textit{adiabatic} in the sense that the four-dimensional expansion can be fast relative to expansion of the extra dimension, which can be assumed to be quasi-static. A key feature of our setup is that, for both constant and time-varying brane energy densities, the 3D branes are, generally, expanding.
One may then wonder whether a standard Friedmann expansion law may be obtained, given that a phenomenologically nonviable, non-standard expansion law in RS~\cite{Binetruy:1999ut,Binetruy:1999hy,Csaki:1999jh,Cline:1999ts,Chung:1999zs} is only cured after stabilizing the radion \cite{Csaki:1999mp,Kanti:1999sz} (see e.g.\ the discussion in \cite[Sec.~8]{Csaki:2000zn} and references therein). We will show that in our scenario the Hubble law receives corrections for energy densities larger than the 5D Planck mass $M$,
but takes its standard form for all cosmologically relevant cases, in particular at late times. Nonetheless we will see that a fine tuning is necessary to obtain realistic values of the absolute expansion rate at each cosmological epoch (say today, at CMB decoupling or during big bang nucleosynthesis(BBN)) and that this fine tuning will be as severe as the standard tuning necessary to obtain the splitting between the cosmological constant and Planck mass. Hence, our model does not give any improvement in the fine tuning, but adds a geometrical picture that may be of use for future approaches. We remark that the natural runaway and expansion of the extra dimension would drive our solution towards the fine tuned point.

Naively, one would expect the presence of a tower of excitable KK modes with a discrete spectrum that is characteristic for the type of warping solution, see e.g.~\cite{Choi:2017ncj}.
Those excited modes would be problematic here, as they could allow information exchange across the horizon. Hence, we analytically compute the spectrum of graviton modes for our solution
and show that they are determined by a modified P\"oschl-Teller potential that possesses exactly two solutions corresponding to graviton states with discrete masses. In addition to the usual massless graviton, 
our scenario predicts an extra massive spin-2 graviton mode with mass $m_1=\sqrt{2}H_0\approx2\times10^{-33}\,\mathrm{eV}$. This mass is understood on top of the usual de Sitter-space offset in mass, $\delta m = \sqrt{2}H_0$, 
which affects both, massless and massive mode (see \cite{Dufaux:2004qs} and references therein e.g.~\cite{Deser:2001wx,Frolov:2002qm}). In addition, we also find a solution in the form of a continuum of massive graviton modes, which is however unphysical because it does not fulfill unitary boundary conditions at the horizon.
The extra massive but light graviton is a unique prediction of this scenario that could be tested for in the future~\cite{deRham:2016nuf,ParticleDataGroup:2022pth}.

As 1999 is almost a quarter of a century ago, we aim to keep the discussion pedagogical and self contained; the paper is organized as follows. 
We start by giving the general action, brane boundary conditions, and Einstein equations, then discuss general solutions and the adiabatic approximation.
Sec.~\ref{sec:NKKK} contains the detailed discussion of the cosmological constant dominated scenario including the derivation of the perceived 4D Planck mass and 4D cosmological constant, 
as well as a comparison of our horizon scenario to the known case of AdS-Schwarzschild. In Sec.~\ref{sec:KKstates} we derive the spectrum of graviton states, the unitary boundary conditions on the horizon, and the effective 4D Newtonian potential. Here we also discuss ways in which the resulting modifications to the standard Newtonian potential might be tested. Finally, in Sec.~\ref{sec:cosmo} we discuss the general case of time-dependent energy densities and pressures on the branes, as well as the hurdles faced in turning this model into a realistic cosmology. In the appendices we briefly review details about how the hierarchy problem is addressed and discuss the normalization of the 4D graviton states.

\section{General considerations}\label{sec:general}
\subsection{Metric and Einstein equations}
We consider a $4+1$ dimensional AdS spacetime\footnote{%
We employ the metric signature $\{+1,\, -1,\,-1,\cdots\}$, Riemann sign convention $\tensor{\mathcal{R}}{_a_b_c^d}=\partial_a\tensor{\Gamma}{^d_b_c}+\cdots$, and denote spacetime indices as $a,b,... \in \{0, 1, 2, 3, 5\}$, $\mu,\nu,...\in\{0, 1, 2, 3\}$, and  $i,j,...\in\{1, 2, 3\}$. Dots and primes denote derivatives w.r.t.\ time and the fifth dimensional coordinate, respectively.} with the line element
\begin{align}
\dd s^2 = g_{ab}\dd x^a\dd x^b = g_{\mu\nu}\dd x^\mu \dd x^\nu + g_{55}\dd z^2 \,,
\end{align}
where the coordinate $x^5=z\in\left\{-\pi R,\pi R\right\}$ corresponds to a compact extra dimension that is symmetric under $z \rightarrow -z$ ($S_1/\mathbbm{Z}_2$ orbifold symmetry). This extra dimension is bounded by two three-dimensional branes located at $z=0,\pi R$ that house all matter content, while the bulk region of the spacetime is assumed to be strictly empty (incl.~the absence of any stabilizing field) and contains only a cosmological constant $\Lambda<0$. We will always assume that our slice of four-dimensional spacetime is located at $z=0$ and call this the ``visible'' brane, while the brane at $z=\pi R$ will be referred to as the ``hidden'' brane. The most general action describing the physics in this geometry is given by
\begin{align} \label{action}
S = \int\dd^4x\,\dd z\,\sqrt{g}\,\left[\frac{M^3}{2}\mathcal{R} - \Lambda - \frac{1}{\sqrt{|g_{55}|}}\big(\Lag_0\,\delta(z) + \Lag_\pi\,\delta(z - \pi R)\big)\right] \,,
\end{align}
where $M$ is the fundamental 5D Planck mass and $\mathcal{R}=g^{ab}\mathcal{R}_{ab}$ is the 5D Ricci scalar, while $\Lag_0$ and $\Lag_\pi$ are arbitrary 4D Lagrangians on the visible and hidden branes.

The 5D Einstein equations obtained from this action take the same form as in 4D, 
\begin{equation}\label{eq:einstein}
\mathcal{G}_{ab} = \mathcal{R}_{ab} - \frac12 g_{ab}\mathcal{R} =  M^{-3}\mathcal{T}_{ab} \,,
\end{equation}
and we assume that the energy-momentum (EM) tensor may be written as
\begin{align} \label{EMT}
\mathcal{T}_a{}^b = -\Lambda\,\delta_a{}^b - 6M^3\,\mathrm{diag}\big[\mathcal{J}_t(t,z), \mathcal{J}_s(t,z), \mathcal{J}_s(t,z), \mathcal{J}_s(t,z), 0\big] \,,
\end{align}
where $\mathcal{J}_{t,s}$ are the temporal and spatial source terms of the 4D branes given by
\begin{align}
&\mathcal{J}_t(t,z) := \frac{1}{6M^3\Bf(t,z)}\Big\{\delta(z)\big[\lambda_0 + \rho_0(t)\big] + \delta(z - \pi R)\big[\lambda_\pi + \rho_\pi(t)\big]\Big\} \,, \return
&\mathcal{J}_s(t,z) := \frac{1}{6M^3\Bf(t,z)}\Big\{\delta(z)\big[\lambda_0 - p_0(t)\big] + \delta(z - \pi R)\big[\lambda_\pi - p_\pi(t)\big]\Big\} \,.
\end{align}
Here, $\lambda_{0,\pi}$ are constant energy densities or ``brane tensions'' (of dimension 4), while $\rho_{0,\pi}$ and $p_{0,\pi}$ are the time-dependent energy densities and pressures of perfect fluids confined to the branes. Just as in standard 4D cosmology, $p$ and $\rho$ are, in general, time-dependent and related by the equation of state $p = w \rho$, where the parameter $w$ takes different values depending on the cosmological era under consideration. In the familiar way, $w=1/3$ corresponds to radiation domination, $w = 0$ to matter domination, and $w=-1$ to epochs dominated by cosmological constants such as inflation or, to leading order approximation, today. For the case $w=-1$ the energy densities $\rho_{0,\pi}$ are constants that simply add to $\lambda_{0,\pi}$. We will see that we can, to an excellent approximation, in all epochs assume that the dynamical quantities $\rho_{0,\pi}$ and $p_{0,\pi}$ are much smaller than their associated brane tensions $\lambda_{0,\pi}$ and still achieve matter or radiation dominated expansion. That is, one should \textit{not} think of $\lambda_{0,\pi}$ as traditional 4D cosmological constants as their effects are largely counterbalanced by 5D quantities in the expansion laws.

For a realistic cosmology, assuming a flat and isotropic 4D spacetime, the most general metric can be written as
\begin{align} \label{genmet}
\mathrm{d}s^2 = \Nf^2(t,z)\,\dd t^2 - \Af^2(t,z)\,\delta_{ij}\dd x^i \dd x^j - \Bf^2(t,z)\, \dd z^2 \,,
\end{align}
where $\Nf(t,z)$, $\Af(t,z)$, and $\Bf(t,z)$ are general functions whose precise form must be determined from the Einstein equations (see also e.g.\ \cite{Binetruy:1999ut,Kim:1999ja})
\begin{align}
(00): \qquad&\frac{1}{\Nf^2}\left(\frac{\dot{\Af}^2}{\Af^2} + \frac{\dot{\Af}\,\dot{\Bf}}{\Af\,\Bf}\right) - \frac{1}{\Bf^2}\left(\frac{\Af''}{\Af} + \frac{\Af'^{\,2}}{\Af^2} - \frac{\Af'\Bf'}{\Af\,\Bf}\right) = - 2\left(\mu^2-\mathcal{J}_t\right)\;, \label{eq:einstein00} \return
(ii): \qquad&\frac{1}{\Nf^2}\left(\frac{\ddot\Af}{\Af} + \frac{\ddot\Bf}{2\Bf} + \frac{\dot{\Af}^2}{2\Af^2} - \frac{\dot{\Nf}\,\dot{\Af}}{\Nf\,\Af} - \frac{\dot{\Nf}\,\dot{\Bf}}{2\Nf\Bf} + \frac{\dot{\Af}\,\dot{\Bf}}{\Af\,\Bf}\right) \nreturn
&- \frac{1}{\Bf^2}\left(\frac{\Nf''}{2\Nf} + \frac{\Af''}{\Af} + \frac{\Af'^{\,2}}{2\Af^2} + \frac{\Nf'\Af'}{\Nf\,\Af} - \frac{\Nf'\Bf'}{2\,\Nf\,\Bf} - \frac{\Af'\Bf'}{\Af\,\Bf}\right) = - 3\left(\mu^2-\mathcal{J}_s\right)\;, \label{eq:einsteinii} \return
(55): \qquad&\frac{1}{\Nf^2}\left(\frac{\ddot\Af}{\Af} + \frac{\dot{\Af}^2}{\Af^2} - \frac{\dot{\Nf}\,\dot{\Af}}{\Nf\,\Af}\right) - \frac{1}{\Bf^2}\left(\frac{\Af'^{\,2}}{\Af^2} + \frac{\Nf'\Af'}{\Nf\,\Af}\right) = -2\mu^2\;, \label{eq:einstein55} \return
(05): \qquad&\frac{\Nf'\dot{\Af}}{\Nf\,\Af} - \frac{\dot{\Af'}}{\Af} + \frac{\Af'\dot{\Bf}}{\Af\,\Bf} = 0\;. \label{eq:einstein05}
\end{align}
Here we have also introduced the positive energy scale $\mu$ defined as\footnote{%
Our definition of $\mu$ here differs from the one used in \cite{Im:2017eju} by a factor $2$.}
\begin{align} \label{muDef}
\mu^2 := -\frac{\Lambda}{6M^3} \,,
\end{align}
to serve as a convenient comparison between the negative bulk AdS cosmological constant and the 5D Planck scale $M$.

It is important to note that the delta distributions in the EM tensor \eqref{EMT} are only supported close to the branes at $z=0,\pi R$, which gives rise to discontinuities of the first derivatives of the metric functions at these locations \cite{Israel:1966}. This implies that, in the supported regions, the second derivatives of the metric functions pick up extra contributions according to
\begin{align}
&\mathcal{F}''(t,z) \enskip\rightarrow\enskip  \mathcal{F}''(t,z) + \delta(z)[\mathcal{F}']_0(t) - \delta(z - \pi R)[\mathcal{F}']_{\pi R}(t) \,,
\end{align}
where $\mathcal{F}=\Nf,\Af$ and $[\mathcal{F}']_{z}:=\mathcal{F}'(z^+) - \mathcal{F}'(z^-)$ denotes a jump of the first derivative. One thus finds that the presence of infinitely thin branes in our 5D spacetime introduces the boundary conditions 
\begin{align}
&\frac{\Nf'(t,0)}{\Nf(t,0)\Bf(t,0)} = -\frac{\lambda_0 - (2 + 3w)\rho_0(t)}{6M^3} \,,
&&\frac{\Af'(t,0)}{\Af(t,0)\Bf(t,0)} = -\frac{\lambda_0 + \rho_0(t)}{6M^3} \,, \label{eq:0bc} \\
&\frac{\Nf'(t,\pi R)}{\Nf(t,\pi R)\Bf(t,\pi R)} = \frac{\lambda_\pi - (2 + 3w)\rho_\pi(t)}{6M^3} \,,
&&\frac{\Af'(t,\pi R)}{\Af(t,\pi R)\Bf(t,\pi R)} = \frac{\lambda_\pi + \rho_\pi(t)}{6M^3} \,, \label{eq:pibc}
\end{align}
where we have used that $[\mathcal{F}']_{z}=2\mathcal{F}'(z)$ at the branes due to the imposed orbifold symmetry. One also encounters non-trivial expressions of energy conservation from the zero component of the covariant EM conservation relation $\nabla_b\mathcal{T}_a{}^b = 0$ in the supported regions,
\begin{align}
\dot{\mathcal{J}}_t = 3\big(\mathcal{J}_s - \mathcal{J}_t\big)\frac{\dot{\Af}}{\Af} - \mathcal{J}_t\frac{\dot{\Bf}}{\Bf} \,,
\end{align}
which simplifies to the following when evaluated at $z=0,\pi R$.
\begin{align} \label{econGen}
&\dot{\rho}_{0}(t) = -3(1 + w)\rho_0(t)\frac{\dot{\Af}(t,0)}{\Af(t,0)} &&\dot{\rho}_\pi(t) = -3(1 + w)\rho_\pi(t)\frac{\dot{\Af}(t,\pi R)}{\Af(t,\pi R)}
\end{align}
This ensures energy conservation on the respective branes.

The presence of discontinuities is generally considered to be unphysical for any spacetime and they may in fact be avoided all together by introducing ``thick branes'' \cite{Csaki:2000fc}. However, it is straightforward to show that there is a smooth limit between the thick brane scenario and the delta function parameterization outlined here. In particular, we find that considering thick branes paired with the appropriate Neumann boundary conditions yields exactly the same relations as \eqref{eq:0bc} and \eqref{eq:pibc}, such that regularizing the delta functions by including non-zero brane thickness does not change our results.

\subsection{Runaway solutions and adiabatic approximation}
Before proceeding with a study of explicit solutions to (\ref{eq:einstein00}--\ref{eq:einstein05}), we address the dynamical size of the extra dimension. 
This size corresponds to the radion metric function $\Bf(t,z)$ and is determined by the $(55)$ component of Einstein's equations independently of the details of the stabilization mechanism \cite{Csaki:1999mp,Kanti:1999sz,Kanti:2000rd}. Throughout this work we assume an empty bulk implying that the $(55)$ equation is strictly the one stated in \eqref{eq:einstein55}. This leaves no possibility to introduce a stabilizing potential for the radion (cf.~e.g.~\cite{Goldberger:1999uk,Kachru:2003aw}) and we are, thus, inherently discussing a
runaway solution for the size of the extra dimension. We emphasize that it is nonetheless possible to obtain a realistic phenomenology if we demand that the evolution of the 5th dimension is slow compared to the evolution of the 3D branes. This motivates the requirement of an \textit{adiabatic} evolution subject to the requirement
\begin{align}\label{eq:adiabatic}
\frac{\dot\Bf}{\Bf} \ll \frac{\dot\Af}{\Af} \,.
\end{align}
Even though runaway solutions are traditionally not very popular, the absence of a stable minimum that realizes 4D de Sitter may, in fact, be preferred from the viewpoint of UV completions of our scenario in the context of string theory, as it generally complies with the de Sitter swampland conjecture \cite{Obied:2018sgi}(see also e.g.~\cite{Denef:2018etk}). Furthermore, the traditionally negative sentiment regarding runaway solutions may be based on the (generally false) perception that static solutions\footnote{%
Here, static refers to the expansion of the 5th dimension and not to the absence of temporal evolution (like Hubble expansion) of the 3D branes.} are necessarily unstable with respect to small perturbations. 
To show that runaway solutions can indeed be viable, we next consider an example as a proof of principle demonstration of a runaway solution that has a smooth limit to a static solution and, hence, can evolve 
adiabatically for a non-zero region of parameters. 

A general solution to Eqs.~\eqref{eq:einstein00}-\eqref{eq:einstein05}, together with the boundary conditions \eqref{eq:0bc} and \eqref{eq:pibc}
for time-independent brane energy densities, has been given by Kim and Kim~\cite{Kim:1999ja}. 
Their solution corresponds to the ansatz\footnote{%
Ref.~\cite{Kim:1999ja} gives the solution in conformal time $\tau$ related to $t$ via $\mathrm{d}\tau/\tau=H_0\,\mathrm{d}t$. In the conformal frame the solution reads $\Nf(\tau,z)=\Af(\tau,z)=\left(\mu\tau\right)^{-1}\Bf(\tau,z)$.}
\begin{align}\label{eq:KKAnsatz}
 \Nf(t,z)~&=~\frac{H_0}{c_b\,H_0\,\e^{H_0t}+\mu\,f(z)}\;,&
 \Af(t,z)~&=~\e^{H_0t}\Nf(t,z)\;,& 
 \Bf(t,z)~&=~\frac{\mu}{H_0}\Nf(t,z)\;,&
\end{align}
where $c_b$ and $H_0$ are constants of mass dimension $0$ and $1$, respectively. 
We note that there is another general solution to the Einstein equations that can be obtained from \eqref{eq:KKAnsatz}
by keeping $\Nf(t,z)$ and $\Af(t,z)$ as above but replacing the ansatz for $\Bf(t,z)$ such that 
\begin{align}\label{eq:JKAnsatz}
 \Nf(t,z)~&=~\frac{H_0}{c_b\,H_0\,\e^{H_0t}+\mu\,f(z)}\;,&
 \Af(t,z)~&=~\e^{H_0t}\Nf(t,z)\;,& 
 \Bf(t,z)~&=~\frac{\mu}{H_0} f(z) \Nf(t,z)\;.&
 \raisetag{22pt}
\end{align}
This gives a different valid solution which has to our knowledge not been discussed so far.
In particular, the Einstein equations demand $f(z)=\sinh(\mu z + c_z)$ (with a dimensionless constant $c_z$) for the original Kim and Kim ansatz while our 
new ansatz requires $f(z)=\csch(\mu z + c_z)$. 
Both solutions share the feature that they describe an extra dimension 
that expands (contracts) for $c_b<0$ ($c_b>0$), while it is static for $c_b\equiv0$.
The crucial difference between our new solution and the Kim and Kim solution is that our solution allows for a smooth limit 
$c_b\rightarrow 0$, while the $c_b=0$ case of the Kim and Kim solution is an isolated point that differs from the solution reached by
taking the limit $c_b\rightarrow 0$. At the exact point $c_b\equiv0$, both solutions coincide up to a coordinate transformation, i.e.\ they merge to a 
single solution corresponding to a static extra dimension. 
This static solution with $c_b\equiv0$ corresponds to a factorization of $\Af(t,z)=a(t)A(z)$ and was independently discovered 
by Nihei~\cite{Nihei:1999mt} and Kaloper~\cite{Kaloper:1999sm}, and has also been discussed more recently in~\cite{Im:2017eju}.
We will soon come back to discuss the static solution in detail, as it corresponds to the $0$th-order approximation for the adiabatic case.

It is important to have a solution such as \eqref{eq:JKAnsatz} that behaves smoothly in the limit $c_b\rightarrow 0$ 
because it allows us to control the evolution of the size of the 5th dimension relative to the 3D branes
adiabatically. In particular, we find that demanding adiabaticity for the solution \eqref{eq:JKAnsatz} corresponds to the requirement
\begin{equation}
 c_b~\ll~\frac{\mu}{H_0}\e^{-H_0 t-\mu \pi R}\;.
\end{equation}
While this shows that $c_b$ must be exponentially suppressed, it also demonstrates the 
existence of a region of parameter space with adiabatic behavior, which is the principle benefit of a solution 
that smoothly connects to the limit $c_b\rightarrow 0$. 
Whether or not such smallness of $c_b$ implies an ``unnatural'' fine tuning is irrelevant for our proof of existence of a well behaved runaway solution.
In any case, the question of fine tuning may only be decided in possible UV completions of our model. 

This shows the existence of runaway solutions consistent with the adiabatic approximation \eqref{eq:adiabatic} for a finite, non-zero 
region of parameter space. We will not further pursue the exact solution \eqref{eq:JKAnsatz}, but focus in the following on the adiabatic
approximation in the next sections. In Sec.~\ref{sec:cosmo} we will discuss the more general case of time-dependent brane energy densities.

\section{Branes with constant energy density}\label{sec:NKKK}
With the general setup established, we proceed to solve the Einstein equations (\ref{eq:einstein00}--\ref{eq:einstein05}) for the case of constant (time-independent) energy densities. This simplified case contains most of the essential features of the model and will subsequently allow us to construct solutions for more general cases, including adiabatic evolution of the extra dimension and branes with time-dependent matter distributions.

\subsection{The NKKK metric and warp factor}
The case of brane energy densities dominated by cosmological constants was first considered by Nihei \cite{Nihei:1999mt}, Kaloper \cite{Kaloper:1999sm}, Kim and Kim \cite{Kim:1999ja} (NKKK). In contrast to earlier works, we allow our spacetime to extend beyond the bulk horizon as, to the best of our knowledge, was first considered in~\cite{Kaloper:1999sm} and then rediscovered in~\cite{Im:2017eju}. We will show that continuing the spacetime in this way is justified in the very same way as it is justified to extend 4D spacetime beyond a black hole horizon e.g.\ in the standard Schwarzschild solution. A key feature of our setup is that, even for constant brane energy densities, the 3D branes are, generally, expanding. This is to be contrasted with the classic Randall Sundrum (RS) case~\cite{Randall:1999ee} which is static with respect to spatial expansion in both the 5th dimension and on the branes. We will see that the RS solution is a special case of the more general solution we present here and that RS represents the particular situation where the brane energy densities are fine-tuned against each other so as to halt 3D expansion \cite{Im:2017eju}.

To begin, we will consider the simplest possible metric ansatz that satisfies all of our previously stated assumptions. Given a vacuum energy dominated equation of state $\rho_{0,\pi}=-p_{0,\pi}$, the matching conditions on the branes \eqref{eq:0bc} tell us that $\Nf'/\Nf=\Af'/\Af$ and suggest that these metric functions are separable, i.e.\  $\Nf(t,z)=n(t)N(z)$ and $\Af(t,z)=a(t)A(z)$, 
which in turn implies that $N(z)=A(z)$ \cite{Lesgourgues:2000tj}. We also assume a time-independent radion $\Bf(t,z)=B(z)$ to satisfy the adiabatic requirement \eqref{eq:adiabatic}. Finally, after redefinitions of the $t$ and $z$ coordinates to absorb the arbitrary functions $n(t)$ and $B(z)$, we arrive at the ansatz
\begin{align} \label{NKKKmet}
\mathrm{d}s^2 = g_{ab}\,\dd x^a \dd x^b = A^2(z)\Big(\dd t^2 - a^2(t)\delta_{ij}\dd x^i \dd x^j\Big) - \dd z^2 \,.
\end{align}
We will refer to this as the ``NKKK metric'', as opposed to the ``general metric'' \eqref{genmet}. Here, $a(t)$ can be associated with the usual FLRW scale factor that describes spatial expansion of the 3D branes in time, leading us to also define the usual Hubble function on the visible brane as
\begin{align}
\mathcal{H}_0(t) := \frac{\dot{a}}{a} \,.
\end{align}
For the NKKK case we will see that $\mathcal{H}_0=const.$, but the same definition also holds for the more general cases below.
Note that $\mathcal{H}_0(t)$ corresponds to the physical expansion rate of a three-dimensional slice of space only at the five-dimensional point $z_0$ where $A(z_0)=1$. The proper physical Hubble rate at a different slice of four-dimensional spacetime, say at $z=z_1$, is given by $\mathcal{H}_1(t)=\mathcal{H}_0(t)/A(z_1)$, as is straightforwardly derived from a redefinition of the time coordinate $A^2(z_1)\dd t^2=\dd \tau^2$. Without loss of generality, we assume our visible brane to be at the origin $z_0=0$ and choose coordinates such that $A(z_0)=1$.

To find a solution for $A(z)$ we consider the Einstein equations (\ref{eq:einstein00}--\ref{eq:einstein05}) in the bulk where $z \neq 0, \pi R$. For the metric \eqref{NKKKmet}, these reduce to
\begin{align}
(00): \qquad&\frac{A''}{A} + \frac{A'^{\,2}}{A^2} - \frac{\mathcal{H}_0^2}{A^2} = 2\mu^2\;, \label{Ec00} \return
(ii): \qquad&\frac{A''}{A} + \frac{A'^{\,2}}{A^2} - \frac{\mathcal{H}_0^2}{A^2} - \frac{2\dot{\mathcal{H}}_0}{3A^2} = 2\mu^2\;, \label{Ecii} \return
(55): \qquad&\frac{A'^{\,2}}{A^2} - \frac{\mathcal{H}_0^2}{A^2} - \frac{\dot{\mathcal{H}}_0}{2A^2} = \mu^2\;. \label{Ec55}
\end{align}
The $(05)$ equation is trivially fulfilled, but after combining the $(00)$ and $(ii)$ equations one finds $\dot{\mathcal{H}}_0(t)=0$ and we will thus use $\mathcal{H}_0(t)=H_0$, where $H_0=const.$ for the rest of this paper. With this, the $(00)$ and $(ii)$ equations become degenerate. For an AdS spacetime that contains nothing other than the cosmological constant in the bulk, the general solution is given by~(see~e.g.~\cite{Binetruy:1999hy,Im:2017eju})
\begin{align}\label{eq:general}
A^2(z) = C_1\,e^{-2\mu z} + C_2\,e^{2\mu z} - \frac{H_0^2}{2\mu^2} \,,
\end{align}
where $C_1$ and $C_2$ are arbitrary constants.\footnote{%
This general solution contains the Randall-Sundrum solution as a static case where $H_0=0$, $C_{2}=0$, and the brane tensions are very finely tuned \cite{Randall:1999ee}.
We refer to~\cite{Im:2017eju} for a more detailed discussion of this and other limits that can be obtained from the general solution.}
Solving the $(55)$ equation, we can further eliminate one of the arbitrary constants in \eqref{eq:general} and fix the other with the normalization requirement $A^2(0)=1$, 
guiding us to the warp factor
\begin{align} \label{Awarp}
&A^2(z) = \frac{H_0^2}{\mu^2}\sinh^2(\mu z - c_z)\;,\qquad\text{where}\qquad c_z := \mathrm{arcsinh}\left(\frac{\mu}{H_0}\right)\;.
\end{align}
Note that there is a location in the extra dimension, $z= c_z/\mu\equiv \pi R_0$, where the warp factor vanishes.
In the literature such points are sometimes referred to as (or in some cases confused with) a ``naked singularity''. 
In our specific case, we will refer to this point as a horizon, drawing the obvious parallel to a Schwarzschild horizon
for reasons that will become clear in the following section.

\subsection{Nature of bulk horizon contrasted with AdS-Schwarzschild}
The fact that the NKKK metric exhibits a zero at $z=\pi R_0$ warrants further investigation. Naturally, one may wonder whether the extra dimension should be regarded as ending at the horizon~\cite{Kaloper:1999sm,Langlois:2000ns} (see also e.g.~\cite{Arkani-Hamed:1998cxo,Gomez:2000bu}).
However, similar to how the gravitational effect of a black hole is felt through its event horizon, we will demonstrate that also the gravitational influence of the hidden brane has important consequences for physics on the visible brane. 
We will now show that in our case, in contrast to previously discussed situations in the literature, there is no singularity at the horizon and thus no problem with extending the space time past the horizon. In subsequent sections, we will also show that unitary boundary conditions at the horizon can be imposed consistently,
implying that causality is preserved and no information can be transported across the horizon.

To simplify the discussion of the horizon, we may focus on the $(t,z)$ plane (every point in this space should be regarded as having a 3D space attached). We also use conventions where time is measured in units of $H_0$ 
and space in units of $\mu$, flip the sign of the $z$ coordinate, and shift the coordinates so that the zero of the metric (the horizon) is located at $\tilde{z}=0$. The resulting 2D metric is then given by
\begin{align}
\dd s^2 = \sinh^2(\tilde{z})\dd t^2 - \dd \tilde{z}^2 \,.
\end{align}
As is commonly done in GR to show that the apparent singularity at the Schwarzschild radius is only a coordinate singularity, we now transform to Kruskal-Szekeres coordinates $\{U,V\}$ with $-UV=\tanh^2(\tilde{z}/2)$ so that our simplified metric is given by
\begin{align}
\dd s^2 = 4\cosh^4\left(\frac{\tilde{z}}{2}\right)\dd U\dd V \,.
\end{align}
This metric is obviously regular and non-vanishing everywhere, in particular at the horizon $(\tilde{z}=0)$, indicating that any apparent singularity at the zero of the metric warping is a coordinate singularity only. Hence, we see that there is merely a causal horizon at $z=\pi R_0$ and not a physical singularity, meaning that there is no need to end the spacetime at this point. 

It is also instructive to compare the present NKKK metric \eqref{NKKKmet} to the anti de Sitter-Schwarzschild (AdS-S) solution, another frequently studied 5D metric with a horizon in the bulk (see~\cite{Witten:1998zw,Horowitz:1998ha,Birmingham:1998nr,Chamblin:1999by,Kraus:1999it,Gubser:1999vj} and the many papers citing them, e.g.~\cite{Hebecker:2001nv,Langlois:2002ke,Langlois:2003zb}).
The metric of AdS-S in our language is given by~(see e.g.~\cite{Kraus:1999it})
\begin{align} \label{AdSSmet}
&\dd s^2 = F^2(z)\dd t^2 - z^2\delta_{ij}\dd x^i\dd x^j - F^{-2}(z)\dd z^2 \,,\quad\text{where}\quad F^2(z) = \mu^2z^2\left(1 - \frac{\pi^4R_0^4}{z^4}\right) \,.
\end{align}
This metric is also a solution to the bulk 5D Einstein equations in AdS space and arises naturally in the study of the AdS/CFT correspondence~\cite{Maldacena:1997re}. 
Both the AdS-S metric and the NKKK metric have event horizons in the bulk and hence, one may wonder if some intuition gained from one case may apply to the other. However, the presence of a metric zero is essentially where the similarities between both solutions end.
While the AdS-S metric is a solution in the presence of a single Planck brane with an unbounded extra dimension \`a la RS2, 
the NKKK metric is an RS1-type solution for a compact extra dimension bounded by both Planck and TeV branes. Moreover, while AdS-S exhibits a physical singularity in the bulk, NKKK only possesses a coordinate singularity.
This fact may be seen clearly by comparing the Kretschmann scalars calculated from each metric,
\begin{align}
&\text{AdS-S:}\quad\mathcal{R}_{abcd}\mathcal{R}^{abcd} = 8\mu^4\left[5 + 9\left(\frac{\pi R_0}{z}\right)^8\right] \,, \qquad  &&\text{NKKK:}\quad\mathcal{R}_{abcd}\mathcal{R}^{abcd} = 40\mu^4 \,.
\end{align}
Clearly, there is a physical singularity at $z=0$ in the AdS-S case, while this scalar curvature takes a constant value for all $z$ in the NKKK case. The two solutions thus describe qualitatively different physical situations.

Furthermore, the Hawking-Unruh temperature of the horizon in the AdS-S case can be interpreted as the temperature of the dual 4D CFT.
By contrast, the temperature of the horizon in the NKKK case corresponds to the rate of 4D Hubble expansion from the point of view of a 4D observer. 
That is, for the case of strictly constant energy densities, the horizon temperature is interpreted as a vacuum energy from the 4D point of view.

We thus see that the presence of the horizon in our case does not necessitate that we truncate the extra dimension or give rise to any pathological behavior, 
motivating us to further explore the phenomenology of this model.

\subsection{Size of the extra dimension and expansion of branes} \label{sec:horizon}
The infinitely thin branes in our setup give rise to discontinuities at each brane, which in turn yield conditions that restrict the brane tensions, 
size of the extra dimension, as well as the expansion rates of the branes, with respect to the bulk fundamental constants. 
Plugging the solution \eqref{Awarp} into the boundary conditions \eqref{eq:0bc}, \eqref{eq:pibc} yields the relations
\begin{align} \label{ABCs}
&-\frac{A'(0)}{A(0)} = \mu\coth(c_z) = \frac{\lambda_0}{6M^3}\;,& &\frac{A'(\pi R)}{A(\pi R)} = \mu\coth(\pi\mu R - c_z) = \frac{\lambda_\pi}{6M^3}\;.
\end{align}
We introduce the parameters 
\begin{equation}
\sigma_{0}:=\frac{\lambda_{0}}{6\mu M^3}=\frac{\lambda_{0}}{\sqrt{-6\Lambda M^3}}\;\qquad\text{and}\qquad\sigma_{\pi}:=\frac{\lambda_{\pi}}{6\mu M^3}=\frac{\lambda_{\pi}}{\sqrt{-6\Lambda M^3}}\;, 
\end{equation}
which provide a dimensionless measure of the brane tensions relative to the fundamental scales of the model, namely 
the bulk cosmological constant and 5D Planck mass. 
The boundary conditions \eqref{ABCs} then imply that there are only consistent solutions for $|\sigma_{0,\pi}|\geq1$.

The three-dimensional expansion of each brane i.e.\ the physical Hubble rate on each brane, can be expressed as a function of the brane tensions as\footnote{%
This inverts eq.~\eqref{ABCs} using $\coth(\arcsinh(x))=\sqrt{1+1/x^2}$.
}
\begin{align} \label{HDefs}
&H_0^2 = \mu^2\big(\sigma_0^2 - 1\big) \,, &&H_\pi^2 = \frac{H_0^2}{A^2(\pi R)} = \mu^2\big(\sigma_\pi^2 - 1\big) \,.
\end{align}
The three important distances in the extra dimension can also be expressed in terms of these tensions as
\begin{equation} \label{RsigDefs}
R_0 = \frac{1}{2\pi\mu}\ln\left[\frac{\sigma_0 + 1}{\sigma_0 - 1}\right]\;,\qquad R_\pi = \frac{1}{2\pi\mu}\ln\left[\frac{\sigma_\pi + 1}{\sigma_\pi - 1}\right]\;,
\end{equation}
and 
\begin{equation} \label{RDef}
R\equiv R_0+R_\pi= \frac{1}{2\pi\mu}\ln\left[\frac{(\sigma_0 + 1)(\sigma_\pi + 1)}{(\sigma_0 - 1)(\sigma_\pi - 1)}\right]\;.
\end{equation}
$R$ is the overall size of the extra dimension (i.e.\ proper distance between visible and hidden brane) while $R_0$ ($R_\pi$) is the proper distance 
between the visible (hidden) brane and the horizon located at the zero of the warp function. It will also be convenient to parameterize the size of the 
extra dimension in terms of the dimensionless constant 
\begin{equation}
\gamma:=R/R_0\;,\qquad\text{such that}\qquad \pi R\equiv\gamma\,c_z/\mu\;.
\end{equation}
In the subsequent sections we will focus exclusively on the case that $R>R_0$, more specifically $R>2 R_0$ (i.e.\ $\gamma>2$),
as this will allow us to address the hierarchy problem. This implies that $\sigma_{0}>1$ and $\sigma_{\pi}>1$, 
i.e.\ both brane energy densities are strictly positive. This feature is a consequence of the fact that the metric warp factor $A$ touches zero in the bulk at a 
distance $R_0$ from the visible brane (the location of the bulk horizon), meaning that $A'$ changes sign at the horizon so that $A$ is cusped upwards on both branes.
The situation is depicted in Fig.~\ref{fig:Aplot}. 

We note that both the original RS1 model (for $\sigma_{0}=-\sigma_{\pi}=\pm1$) and the original RS2 model
(for $R_0\rightarrow\infty$) are contained as special static ($H_0=0$) cases of our general solution and we refer to~\cite{Im:2017eju} for a more detailed discussion of these relationships. 
Our parameter region $\sigma_{0,\pi}>1$ is to be contrasted with the parameter region originally discussed in \cite{Kim:1999ja}, where 
$1<\sigma_0<-\sigma_\pi$ such that $R<R_0$. In this case the case the presence of a bulk horizon is avoided, but this necessarily comes at the cost of an unphysical negative energy density on one of the branes, just as in original RS1 model~\cite{Randall:1999ee}.
For completeness, we also note that it would in principle be possible to take $R=R_0$ as a valid solution for the size of the extra dimension, in which case the hidden brane decouples and there are solutions for any value of $\sigma_\pi$ provided that $\sigma_0>1$~\cite{Im:2017eju}.

We stress that our solution is tightly constrained and all parameters are functionally related. We will see that requiring a realistic phenomenology
and a natural explanation of the electroweak hierarchy fixes all free parameters. Intuitively, this may be understood by taking the brane energy densities $\sigma_0$ and $\sigma_\pi$
as free and independent parameters. The corresponding expansion rates of the 4D branes are then fixed via $\eqref{HDefs}$,
which also fixes the distance of each of the branes to the horizon $R_0$ and $R_{\pi}$ and, therefore, the overall size of the 
extra dimension~\eqref{RsigDefs}. One may imagine the two sides separated by the causal horizon as two causally disconnected sides of one universe that evolve independently subject to their respective 4D brane energy densities. The overall size of the extra dimension then simply results as the sum of the distances of the two branes to the horizon, $R=R_0+R_{\pi}$. The fact that the observed strength of 4D gravity 
depends on the overall size of the extra dimension gives a non-local connection between the two sides, at least at the time of the horizon's formation, even though we do not speculate on any specific formation mechanism here. 
Requiring a realistic 4D expansion rate $H_0$ fixes $\sigma_0$ (and thus the distance to the horizon, $R_0$), while accommodating for a realistic 4D Planck mass $M_p$ fixes the overall size of the extra dimension $R$ (and thus also $\sigma_\pi$). 
We will discuss the details of this picture in the following section.

\subsection{4D Hierarchy problem and cosmological constant}\label{sec:HPandCC}
Let us now discuss how the hierarchy and cosmological constant problem are addressed in this setup. 
In order to avoid the hierarchy problem, all fundamental scales in our model should be of approximately the same order. 
In particular, this implies $M\sim|\Lambda|^{1/5}\sim\mu\sim v_{\mathrm{EW}}$ where $v_{\mathrm{EW}}$ is the perceived
fundamental electroweak (EW) breaking scale (see App.~\ref{sec:HP} for specific details). 

The 4D Planck mass and cosmological constant are emergent scales that may be derived from the effective 4D action obtained by integrating out the extra dimension (performing the integral over $z$) in the complete 5D action~\eqref{action}, in the same way as in the original RS model.\footnote{%
One may consider whether this integral should be taken over the entire extra dimension or if it should be bounded by the horizon. The RS solution to the hierarchy problem is sometimes obfuscated with apparently local statements about the effective 4D strength of gravity being weakened by a flux of gravitons ``leaking'' into the extra dimension, which may lead one to assert that only causally connected regions should contribute to the weakening of 4D effective gravity. However, we stress that the perceived strength of 4D gravity is by its very definition a global property of each 4D slice. The situation here is perfectly analogous to that of a standard Schwarzschild black hole. A test mass orbiting the black hole will naturally be subject to the gravitational influence of the mass of the central singularity despite the presence of the event horizon -- if one computed an integral of the density of the black hole to determine the force on the test particle, they would get zero if they bounded that integral at the event horizon. We thus assert that the integral over $z$ which determines the effective Planck mass must be evaluated over the entire extra dimension and across the horizon, especially since we are agnostic about the history and mechanism by which the 5D horizon has formed.}
The separability of the 5D metric \eqref{NKKKmet} allows us to write
\begin{equation}\label{eq:dimred}
g_{\mu\nu}(x^a) = A^2(z)\hat{g}_{\mu\nu}(x^\mu)\,,\quad \sqrt{g(x^a)} = A^4(z)\sqrt{-\hat{g}(x^\mu)}\,,\quad \mathcal{R}(x^a) = A^{-2}(z)\hat{\mathcal{R}}(x^\mu) + f(z)\,,
\end{equation}
where $f(z)$ is a function of $z$ only,\footnote{%
For the metric \eqref{NKKKmet}, $f(z)=-12A'^2/A^2-8A''/A$.
} and hats indicate quantities defined with respect to the effective 4D FLRW metric
\begin{align} \label{ghat}
\hat{g}_{\mu\nu}\,\dd x^\mu \dd x^\nu = \dd t^2 - a^2(t)\delta_{ij}\dd x^i \dd x^j \,.
\end{align}
The effective 4D action of interest is simply the standard Einstein-Hilbert action 
\begin{align} \label{eq:EHaction}
S_\text{EH} = \int\dd^4x\sqrt{-\hat{g}}\left(\frac{M_p^2}{2}\hat{\mathcal{R}} - \hat{\Lambda}\right) \,.
\end{align}
Inserting \eqref{eq:dimred} into \eqref{action}, we can compute the prefactor of the Ricci scalar term
of the 4D Einstein-Hilbert action, resulting in
\begin{equation} \label{MpM}
\begin{split}
\frac{M_p^2}{M^3} &= \int_{0}^{2\pi R}\dd z A^2(z) = 2 \int_{0}^{\frac{\gamma c_z}{\mu}}\dd z A^2(z) \\
                  &= \frac{H_0^2}{2\mu^3}\Big\{\sinh\left[2c_z(\gamma - 1)\right] + \sinh(2c_z) - 2c_z\gamma\Big\} 
\approx \frac{1}{\mu}\left(\frac{2\mu}{H_0}\right)^{2\gamma - 4} \,,
\end{split}
\end{equation}
where the last approximation is valid only for $\mu \gg H_0$ and $\gamma>2$.\footnote{%
The subleading terms $M_p^2/M^3=\mu^{-1}[(2\mu/H_0)^{2\gamma-4}+1+\mathcal{O}\{(2\mu/H_0)^{2\gamma-6}\}]$ are 
completely negligible for all our purposes. Nonetheless, we display them here to highlight the fact that 
the $\mu$ dependence of $M_p$ drops at $\gamma=5/2$ only to leading order. 
For the symmetry enhanced point $\gamma=2$ the exact result is given by $M_p^2/M^3=\mu^{-1}$.} 

We can also compute the perceived 4D cosmological constant on the visible brane, $\hat\Lambda$, starting 
from the expansion law \eqref{HDefs}. Since $H_0\ll\mu$, we parametrize $\sigma_0=1+\delta\sigma_0$ where $0\leq\delta\sigma_0\ll1$ 
and the perturbation is given precisely by 
\begin{equation}
\delta\sigma_0\equiv\frac{\lambda_0}{6M^3\mu}-1\equiv\frac{\delta\lambda_0}{6M^3\mu}\;.
\end{equation}
To leading order in $\delta\sigma_0$,
\begin{equation}
\frac{\hat\Lambda}{3M_p^2}~\stackrel{!}{=}~H_0^2~\approx~2\mu^2\delta\sigma_0~=~\frac{\mu\,\delta\lambda_0}{3M^3}~\approx~\frac{\delta\lambda_0\left(\frac{2\mu}{H_0}\right)^{2\gamma-4}}{3M_p^2}\;,
\end{equation}
where the first step is just the regular 4D Friedmann expansion law following from \eqref{eq:EHaction}, the second step expands \eqref{HDefs} to leading order,
and for the last step we have used \eqref{MpM}.
Hence, we can read off
\begin{equation}
 \hat\Lambda~\approx~\delta\lambda_0\left(\frac{2\mu}{H_0}\right)^{2\gamma-4}\;.
\end{equation}
We will also see that $\delta\sigma_\pi\equiv\sigma_\pi-1\ll1$ is required, and in this limit one can derive 
\begin{equation}
M_p^2~\approx~\frac{M^3}{\mu}\frac{\delta\sigma_0}{\delta\sigma_\pi}\;.
\end{equation}
We may also derive equivalent expressions that, instead of depending on $\delta\sigma_{0,\pi}$, express our results as functions of the 
proper size of the extra dimension:
\begin{align}
 H_0 &~\approx~ 2\mu\,\e^{-\pi R_0 \mu}\;,&  &H_\pi ~\approx~ 2\mu\,\e^{-\pi R_\pi \mu}\;,& \\ \label{eq:Mpl}
 M_p^2 &~\approx~ \frac{M^3}{\mu}\,\e^{2\pi\mu\left(R_\pi-R_0\right)}\;,& \\
 \hat\Lambda &~\approx~ 12\,M^3\mu\,\e^{-2\pi\mu\left(2R_0-R_\pi\right)}\;.&
\end{align}

To put bounds on the parameters of our model, we require that the correct (reduced) Planck mass, $M_p=2.4\times10^{18}\,\mathrm{GeV}$, and the observationally determined value of the current expansion rate, $H_0=1.5\times10^{-33}\,\mathrm{eV}$, are reproduced. For the fundamental 5D scales, we adopt the universal benchmark value $M\sim|\Lambda|^{1/5}\sim\mu\sim\mathrm{TeV}$ and consider the hierarchy problem to be solved if all these 
fundamental parameters vary within $\pm2$ orders of magnitude around the $\mathrm{TeV}$ scale, which restricts $\gamma$ to the interval $\gamma\in[2.25,2.45]$. 
Finally, the observed Hubble rate requires $\delta\sigma_0\sim10^{-88}$ and one finds that every other unmentioned quantity is fixed in terms of these parameters, for example, $\delta\sigma_\pi\sim10^{-118}$.
The actual tuning required to achieve realistic parameters is $\delta\lambda_0/\lambda_0\sim10^{-88}$ and $\delta\sigma_\pi/\delta\sigma_0\sim10^{-30}$,
and hence, in total, does not represent an improvement over the standard cosmological constant problem. However, the smallness of parameters may at least be understood in a natural way as each of the points $\sigma_0=1$, $\sigma_\pi=1$, and $\gamma=2$ is special; the first two restore Minkowski space on the respective brane while the latter introduces a $\mathbbm{Z}_2$ exchange symmetry of the branes.

Interestingly, for realistic parameters the proper distance of the visible brane to the horizon comes out as $\mu\pi R_0\approx 104$, while the proper distance 
between visible and hidden brane, i.e. the overall size of the extra dimension results in $\mu\pi R\approx243$. 
We find it quite remarkable that the two most widely separated scales in Nature, namely the cosmological constant and the Planck mass, correspond in our picture to two distance scales in one extra dimension that differ only by a factor $\gamma\sim2.34$.
As a final note, we also point out that the physical size of the extra dimension (under the assumption that $\mu\sim\mathrm{TeV}$) can be computed as $R\approx10^{-17}\,\mathrm{m}$.

\section{4D Graviton states} \label{sec:KKstates}
\subsection{Equations of motion and mass eigenstates}
We now move on to a discussion of the gravitational perturbations in the cosmological constant dominated epoch that is governed by the NKKK metric \eqref{NKKKmet} discussed in the previous section. This metric is most relevant for the phenomenology of the present universe which is cosmological constant dominated and its separability in $t$ and $z$ allows for straightforward analytical computations that do not rely on additional approximations. In Sec.~\ref{sec:cosmo} we will discuss the more general case with time varying energy densities on the branes,
which turns out to amount to only a small perturbation to the present case. Hence, we expect the discussion of this section to hold qualitatively and, to first approximation, quantitatively for the more general case as well.

Perturbations of the 5D metric may be decomposed into 4D Kaluza-Klein (KK) modes that represent a spectrum of gravitational perturbations when viewed from the branes. The spin-2 tensor (graviton) and spin-0 scalar (radion) perturbations are of particular importance for establishing the phenomenological viability of the 4D effective theory, whereas the spin-1 vector excitations are generally too massive to be relevant in setups with orbifold symmetry \cite{Randall:1999ee}. As a result of 4D Poincar\'e invariance on the branes, we are guaranteed that the KK tensor zero mode is massless~\cite{Csaki:2000fc}. In addition to the massless mode, we will see that there arises a new light graviton mode with mass below detection threshold and that, together, these modes are responsible for long range 4D gravity. On the other hand, a light dynamical radion mode is tightly constrained by experimental fifth-force constraints and should, therefore, not appear if our model is to be phenomenologically viable. In the RS theory, the radion corresponds to the modulus of the extra-dimensional coordinate and its vacuum expectation value determines the size of the extra dimension (which is usually stabilized via the Goldberger-Wise mechanism \cite{Goldberger:1999uk}). However, in the present model, the size of the extra dimension \eqref{RDef} is fixed by our exact solution to the (55) Einstein equation \eqref{Ec55} and depends on the relative sizes of the fundamental 5D scales and the brane tensions. This precludes the RS radion from appearing as a dynamical DOF in the 4D theory by assumption. This is not to say that our exact solution to the (55) equation necessarily excludes the possibility that the (55) component of the 5D graviton perturbations can appear as a dynamical field in the 4D effective theory. It is still important to investigate this possibility and we will explicitly show that such a radion is non-dynamical in our setup.

We consider graviton perturbations $h_{\mu\nu}(x^\mu,z)$ and radion perturbations $\phi(x^\mu,z)$ in the expansion
\begin{align} \label{pertmet}
g_{ab}\,\dd x^a\dd x^b \to A^2(z)\big(\hat{g}_{\mu\nu} + h_{\mu\nu}\big)\dd x^\mu\dd x^\nu - \big(1 - \phi\big)\dd z^2 \,.
\end{align}
In order to establish equations of motion (EOMs) for these perturbations, we follow a method similar to that of Cs\'aki et al.~\cite{Csaki:2000fc} and begin with a coordinate transformation $z \rightarrow u$ where $u(z)$ solves the differential equation
\begin{align} \label{confDE}
\left(\frac{\dd u}{\dd z}\right)^2 = \frac{1}{A^2(z)} \,.
\end{align}
Using our warp factor \eqref{Awarp}, the explicit solutions to \eqref{confDE} are found to be
\begin{align}\label{eq:uz}
u(z) = \frac{1}{H_0}\left\{c_u \pm \log\left|\coth\left(\frac{\mu z - c_z}{2}\right)\right|\right\}\,,
\end{align}
where $c_u$ is an arbitrary constant of integration that may be set so that the visible brane remains at the origin i.e.\ so that $u(0)=0$. Selecting the ``$+$'' branch of this solution then implies that the horizon lies at $+\infty$ in $u$.\footnote{%
Specifically, we take $c_u=-\log\left|\coth\left(\frac{c_z}{2}\right)\right|$ such that $z\in[0,c_z/\mu)$ corresponds to $u\in[0,\infty)$, noting that the $u$ coordinate chart defined by \eqref{eq:uz} can not cover the whole extra dimension but only the region up to the horizon. We speculate that working in conformal coordinates from the start may be one reason why our whole solution has been missed or discarded in the past.
It is no problem to work and solve the EOMs in $u$, and later extend the resulting solutions back to $z\in[-\pi R, \pi R]$, making sure they solve the EOMs also for the extended domain.}
Inverting \eqref{eq:uz} and plugging it back into the physical warp factor \eqref{Awarp} gives us the warp factor in conformal coordinates
\begin{align} \label{Au}
A^2\big(z(u)\big) =  \frac{H_0^2}{\mu^2}\csch^2(H_0 u - c_u) \,.
\end{align}
After applying this coordinate transformation to the metric \eqref{NKKKmet}, we may separate the warp factor $A^2\big(z(u)\big)$ as a conformal pre-factor and write
\begin{gather} \label{bgmet}
g_{ab} = A^2\big(z(u)\big)\;\tilde{g}_{ab}\,, \quad\text{where}\quad \tilde{g}_{ab}\,\dd x^a \dd x^b = \dd t^2 - a^2(t)\delta_{ij}\dd x^i\dd x^j - \dd u^2 \,.
\end{gather}
Expressing our theory in terms of perturbations of $\tilde{g}_{ab}$ will allow us to derive an EOM for the KK tensor modes that correspond to the 4D perturbations of interest. After transforming to the $u$-coordinates and using the conformal decomposition \eqref{bgmet}, the bulk action of our theory can be written as 
\begin{align} \label{SConf}
S_\text{bulk} = \int\dd^4x\,\dd u\,\sqrt{\tilde{g}}\left[\frac{M^3A^3}{2}\left(\tilde{\mathcal{R}} + \frac{4\tilde{\nabla}_aA\tilde{\nabla}^aA}{A^2} + \frac{8\tilde{\nabla}_a\tilde{\nabla}^aA}{A}\right) - A^5\Lambda\right] \,,
\end{align}
where indices are raised and lowered with $\tilde{g}_{ab}$, and $\tilde{\nabla}$ denotes the covariant derivative w.r.t.~$\tilde{g}_{ab}$. The 5D metric fluctuations are exposed by perturbing the metric and expanding to second order in the perturbations.
Specifically, we perturb according to
\begin{align} \label{tgexp}
\tilde{g}_{ab} \to \tilde{g}_{ab} + h_{ab} + \Phi_{ab} \,,
\end{align}
with the constraints
\begin{align} \label{hcompdefs}
h_{a5} = h_{5a} = \Phi_{a\mu} = \Phi_{\mu a} = 0 \qquad\text{and}\qquad \Phi_{55} = \phi
\end{align}
fixed on the components, in order to remove all contributions from the (independent) spin-1 perturbations and ensure that we recover \eqref{pertmet} after converting back to $z$-coordinates. We also work in the transverse traceless gauge with the gauge conditions
\begin{align} \label{ttgauge}
&\tilde{\nabla}_bh_a{}^b = 0 \,, &h_a{}^a = 0 \,.
\end{align}

All together, after expanding \eqref{SConf} to second order per \eqref{tgexp}, integrating by parts, and simplifying terms with \eqref{hcompdefs} and \eqref{ttgauge}, the quadratic part of the perturbed action is found to be
\begin{align}
S_\text{pert} = \frac{M^3}{4}\int\dd^4x\,\dd u\,\sqrt{\tilde{g}}\bigg[&A^3\left(\frac12\tilde{\nabla}_ch_{ab}\tilde{\nabla}^ch^{ab} - H_0^2h_{ab}h^{ab}\right) \nonumber\\
&- 3\phi^2\Big(A\tilde{\nabla}_aA\tilde{\nabla}^aA - H_0^2A^3 - \mu^2A^5\Big)\bigg] \,,
\end{align}
where we have also made use of the relations $h^{cd}\tilde{\mathcal{R}}_{acbd}=-H_0^2h_{ab}$ and $\Phi^{cd}\tilde{\mathcal{R}}_{acbd}=0$ after commuting covariant derivatives.

It is readily apparent that the radion has no kinetic term in this action and is thus non-dynamical in the present theory. Recalling \eqref{hcompdefs}, one can see how this occurs as a result of cancellations of the form
\begin{align}
\tilde{\nabla}_c\Phi_{ab}\tilde{\nabla}^c\Phi^{ab} - \tilde{\nabla}_c\Phi_a{}^a\tilde{\nabla}^c\Phi_b{}^b = \tilde{\nabla}_c\phi\tilde{\nabla}^c\phi - \tilde{\nabla}_c\phi\tilde{\nabla}^c\phi = 0
\end{align} 
and similar, whereas the analogous terms for the graviton do not cancel due to its traceless-ness. The radion EOM is thus simply the constraint
\begin{align}
\left[\left(\frac{A'}{A}\right)^2 + \mu^2A^2 + H_0^2\right]\phi = \coth(H_0u - c_u)\phi = 0 \,,
\end{align}
which implies that $\phi=0$. Varying the action with respect to $h_{ab}$ on the other hand, we find the non-trivial EOM
\begin{align} \label{EOMlin}
\left(\tilde{\nabla}_c\tilde{\nabla}^c + \frac{3\tilde{\nabla}_cA\tilde{\nabla}^c}{A} + 2H_0^2\right)h_{ab} = 0 \,.
\end{align}

We obtain an equation for the 4D mass eigenstates from this equation by applying the KK decomposition ansatz 
\begin{align} \label{KKdecomp}
h_{\mu\nu}(x^\mu,u) ~=~\sum_n^\infty \hKK{n}_{\mu\nu}(x^\mu)\;\Psi_n(u) \,, 
\end{align}
which expresses the 5D graviton in terms of effective 4D graviton eigenstates\footnote{%
This decomposition is essentially just a Fourier transformation and in the case that the eigenstates form a continuous basis, this sum is replaced by an integral over $n$.} that are separable in $x^\mu$ and $u$ \cite{Bailin:1984xf}. 
The $\hKK{n}_{\mu\nu}$ respect the 4D Klein-Gordon equation
\begin{align} \label{KGeqn}
\left(\tilde{\nabla}_\rho\tilde{\nabla}^\rho + m_n^2 + 2H_0^2\right)\hKK{n}_{\mu\nu} = 0 \,,
\end{align}
which is derived from the linearized 4D Einstein-Hilbert action 
after including a Fierz-Pauli mass term (in agreement with \cite{Dufaux:2004qs}). 
It is also advantageous to define rescaled $u$-dependent ``wavefunctions'' as
\begin{align} \label{resacling}
\Psi_n(u) = A^{-3/2}(u)\psi_n(u) \,,
\end{align}
which has the effect of removing the $\tilde{\nabla}A\tilde{\nabla}h$ cross term in \eqref{EOMlin}. 
Applying eqs.~\eqref{KKdecomp}, \eqref{KGeqn}, and \eqref{resacling}, we find that the graviton EOM eq.~\eqref{EOMlin} reduces to
\begin{align} \label{EOMh}
\hKK{n}_{\mu\nu}\left\{\eta^{ab}\left[\frac{\partial_a\partial_b\psi_n}{\psi_n} - \frac{3}{2}\left(\frac{\partial_a A\partial_b A}{2A^2} + \frac{\partial_a\partial_b A}{A}\right)\right] - m_n^2\right\} = 0 \,.
\end{align}

Rewriting \eqref{EOMh} in a more familiar form, the initially complicated 5D graviton EOM reduces to a one-dimensional Schr\"odinger equation for $\psi_n(u)$,
\begin{align} \label{Schrod}
&\psi_n''(u) - \left[V(u) - m_n^2\right]\psi_n(u) = 0\,,\qquad\text{with}\qquad V(u) = \frac{3A''}{2A} + \frac{3\big(A'\big)^2}{4A^2}\;.
\end{align}
Despite the fact that we have used an FLRW-style background metric \eqref{bgmet} here, 
this result agrees with the warp-factor-independent analysis presented in \cite{Csaki:2000fc}, 
where the 4D background metric is taken to be flat Minkowski space.
Plugging in the warp factor in the conformal-basis, eq.~\eqref{Au}, the potential in \eqref{Schrod} is found to be
\begin{align} \label{PTpot}
V(u) = \frac{15}{4}\csch^2(H_0u - c_u) + \frac{9}{4}H_0^2 \,,
\end{align}
which fits into the class of ``modified P\"oschl-Teller'' potentials \cite{Poschl:1933zz,Diaz:1999}. These kind of potentials have been known to appear in other extra-dimensional models 
such as~\cite{Brandhuber:1999hb,Herrera-Aguilar:2009azq,Barbosa-Cendejas:2007ucz}, however, our potential differs slightly from these, most obviously in the ``$\csch$'' instead of the usual ``$\sech$'', and these small differences will turn out to have a significant impact on the KK spectrum. 
The appearance of a P\"oschl-Teller-type potential is an attractive feature because it implies that the Schr\"odinger equation \eqref{Schrod} can be solved analytically. The general solutions are given by
\begin{equation} \label{gensolu}
\psi_n(u) = a_n\,P_{3/2}^\alpha\!\left(\coth\left[H_0u - c_u\right] \right) + b_n\,Q_{3/2}^\alpha\!\left( \coth\left[H_0u - c_u\right]\right)\;,
\end{equation}
where $P_{3/2}^\alpha(x)$ and $Q_{3/2}^\alpha(x)$ are associated Legendre functions of the first and second kind of degree $3/2$ and order
\begin{equation}
 \alpha = \sqrt{\frac{9}{4} - \frac{m_n^2}{H_0^2}}\;,
\end{equation}
while $a_n$ and $b_n$ are arbitrary constants of integration to be fixed by boundary conditions.

This is interesting because the Legendre function solutions, for $\alpha\in\Reals$, only result in phenomenologically viable\footnote{%
The Legendre functions in \eqref{gensolu} only have finite series representations for order $\alpha=k/2$ when $k$ is odd and are thus non-normalizable out of this regime. For $\alpha>3/2$ that are members of this set we encounter tachyonic solutions, though these eigenstates are excluded anyway by the unitary boundary conditions that we derive in the next section.} 
wavefunctions if $\alpha\in\{1/2,3/2\}$~\cite{Bateman,Herrera-Aguilar:2009azq}.
This implies the presence of a mass gap between the massless eigenstate corresponding to $(m_0,\alpha=3/2)$ and the excited eigenstate corresponding to $(m_1,\alpha=1/2)$ with
\begin{equation} \label{discmasses}
m_0 = 0 \qquad\text{and}\qquad m_1 = \sqrt{2}H_0 \;.
\end{equation}
There is an additional case to consider, as the general solution \eqref{gensolu} may be normalizable also for purely imaginary $\alpha$, i.e.\ for $m_n^2 > 9/4H_0^2$, in which case it asymptotes to plane waves. Writing $\alpha=i\beta$ where $\beta\in\Reals^+$, we thus find an additional continuum of possible mass states
\begin{align} \label{contmasses}
m_\beta ~=~ \frac{H_0}{2}\sqrt{4\beta^2 + 9}~\geq~ \frac{3}{2}H_0\,.
\end{align}

In total, the potential \eqref{PTpot} thus allows for a massless 4D graviton, a discrete massive state, and a continuum of massive states separated by a mass gap. This kind of general spectrum is expected from the modified P\"oschl-Teller potential, as seen e.g.\ in \cite{Herrera-Aguilar:2009azq,Barbosa-Cendejas:2007ucz}. The present setup differs from earlier works, however, where the mass gap is typically quite large ($m_n$ there is proportional to the inverse brane width), 
while here the mass gap is of order $\Ord(H_0)$. Needless to say, the presence of light graviton states, in particular the continuum of light states, might present a phenomenological issue at first glance. However, as we will discuss next, the continuum solutions are in fact excluded after application of appropriate boundary conditions.

\subsection{Unitary boundary conditions}
There is good precedent in the literature regarding how to choose a suitable set of boundary conditions in the presence of bulk horizon in extra-dimensional models. We follow the original work of Gell-Mann and Zwiebach~\cite{Gell-Mann:1985lxb} (and more recently \cite{Herrera-Aguilar:2009azq,Barbosa-Cendejas:2007ucz}) to select ``unitary boundary conditions'' which guarantee the conservation of 4D energy and momentum such that no conserved 4D quantity is allowed to leak through the horizon. 
These 4D conservation laws correspond to the $x^\mu$ translational isometries of the metric $\tilde{g}_{ab}$, which are generated by the Killing vectors $\xi_{(\mu)}^a$ that satisfy the Killing equation
\begin{align} \label{Killeq}
\tilde{\nabla}^a\xi_{(\mu)}^b + \tilde{\nabla}^b\xi_{(\mu)}^a = 0 \,.
\end{align}
Each of these Killing vectors corresponds to a covariantly conserved current,
\begin{align}
&J_{(\mu)}^a = \xi_{(\mu)}^b\mathcal{T}^a{}_b \,, &\frac{1}{\sqrt{\tilde{g}}}\partial_a\left(\sqrt{\tilde{g}}J_{(\mu)}^a\right) = 0 \,,
\end{align}
where $\mathcal{T}_{ab}$ is the EM tensor derived from the bulk graviton Lagrangian. Unitary boundary conditions are formulated based on the requirement that the flux across the horizon (the transverse component of these currents) vanishes,
\begin{align}
\lim_{u\rightarrow\infty}\sqrt{\tilde{g}}J_{(\mu)}^{\,5} = 0 \,.
\end{align}

From \eqref{bgmet} and \eqref{Killeq}, the 4D translational Killing vectors are found to be
\begin{align}
&\xi_{(0)}^a = \{1,-H_0\,x^1,-H_0\,x^2,-H_0\,x^3,0\} \,, &\xi_{(i)}^a = \delta_i^a \,,
\end{align}
with the associated currents given by
\begin{align}
J_{(\mu)}^{\,5} ~=~ \xi_{(\mu)}^\nu\,\hat{h}^{(n),\rho\sigma}\left(\frac14\tilde{\nabla}_\rho\hKK{n}_{\nu\sigma} - \frac38\tilde{\nabla}_\nu\hKK{n}_{\rho\sigma}\right)A^3\,\tilde{g}^{a5}\,\Psi_n\,\tilde{\nabla}_a\Psi_n \,.
\end{align}
A necessary condition for the conservation of 4D energy and momentum, hence, is given by
\begin{align} \label{uniCon}
\lim_{u\rightarrow\infty}\;A^3\,\Psi_n\,\partial_u\Psi_n = 0 \,.
\end{align}

We can apply this condition to each of the mass eigenstates found in the previous section, being careful to restore the full $u$ dependence of each solution by using \eqref{resacling}. 
For the ground state $(m_0=0,\alpha=3/2$), the general solution results in
\begin{align}\label{eq:Psi0}
\Psi_0(u) = b_0\left\{\cosh\left[H_0u - c_u\right] - \frac19\cosh\left[3\left(H_0u - c_u\right)\right]\right\}+a_0 \,.
\end{align}
Requiring~\eqref{uniCon} then enforces $b_0=0$ to ensure 4D energy-momentum conservation. 
The first excited state $(m_1=\sqrt{2}H_0,\alpha=1/2$) is given by
\begin{align}\label{eq:Psi1} 
\Psi_1(u) = b_1\left\{\cosh\left[2\left(H_0u - c_u\right)\right] + 3\right\} + a_1\cosh\left(H_0u - c_u\right)\,,
\end{align}
and~\eqref{uniCon} similarly requires $b_1=0$.

\begin{figure}[!t!]
\centering
\includegraphics[width=1.0\linewidth]{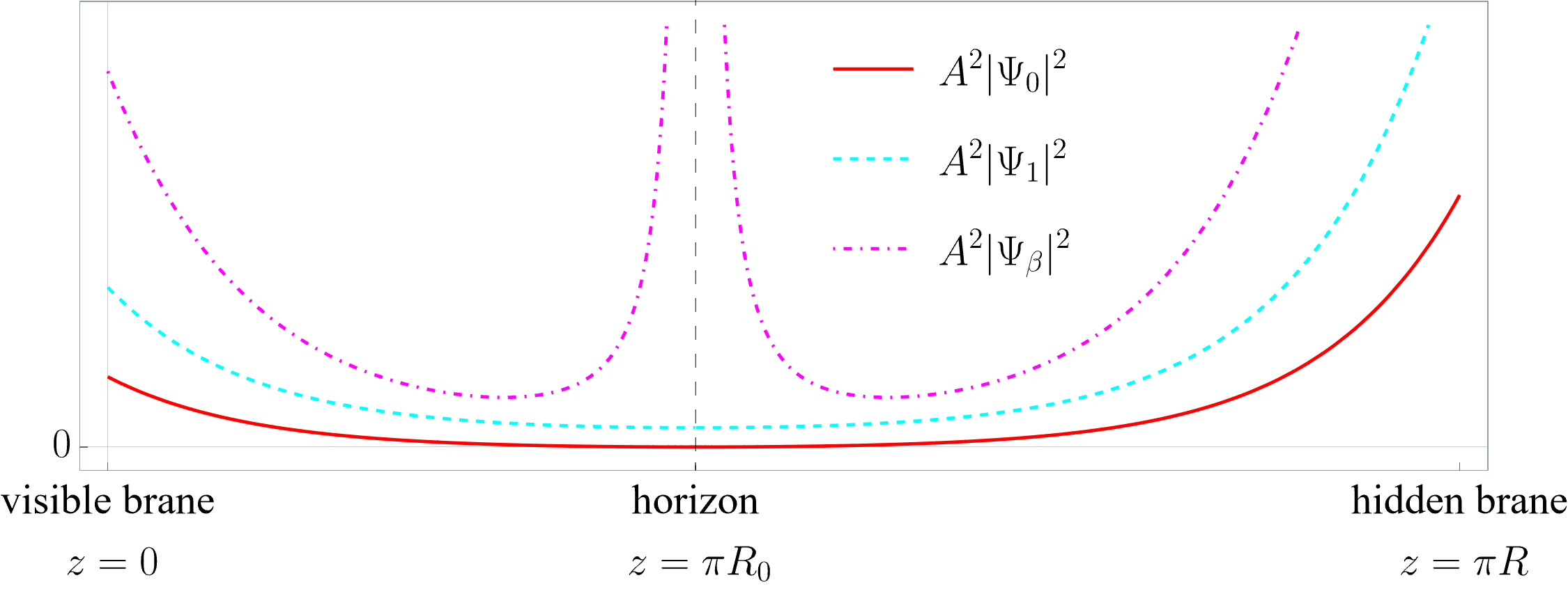}
\caption{\label{fig:yplot}
Graviton profiles along the extra dimension for the massless mode $\Psi_0$, first excited state $\Psi_1$, and the continuum modes $\Psi_\beta$, each absolute squared 
and multiplied by the metric warp factor $A^2$ as they appear in the normalization integral. The continuum modes, $A^2|\Psi_\beta|^2$, diverge at the horizon signaling their unphysical nature.}
\end{figure}

The analysis of the continuum modes $\Psi_\beta$ requires more care, as the Legendre functions in~\eqref{gensolu} do not have a simple series representation for general $\alpha=i\beta$,
in contrast to the case of real and discrete $\alpha$. 
However, to test energy-momentum conservation close to the horizon, it is sufficient to work with asymptotic expansions for $P^{i\beta}_{3/2}$ and $Q^{i\beta}_{3/2}$, see~\cite[ch.\ 10.14]{Bateman}. For $u\rightarrow\infty$, corresponding to $\coth(H_0u-c_u)\rightarrow1$, both Legendre functions become singular, scaling like
\begin{align}
&\lim_{x \rightarrow 1}P_{3/2}^{i\beta}(x)\; \propto\; \lim_{x \rightarrow 1}(x - 1)^{-i\beta/2}& &\text{and}& &\lim_{x \rightarrow 1}Q_{3/2}^{i\beta}(x)\;\propto\;\lim_{x \rightarrow 1}(x - 1)^{i\beta/2}\;.&
\end{align}
The complex exponentials expose oscillatory behavior close to the horizon and the continuum wavefunctions in the asymptotic limit can be represented as
\begin{align}
\Psi_\beta(u) \;\sim\; \sinh^{3/2}\!\left(H_0u - c_u\right)\; \left[a_\beta\,\cos\left(\beta H_0u\right) + b_\beta\,\sin\left(\beta H_0u\right)\right]\;,
\end{align}
with general complex constants $a_\beta$ and $b_\beta$. To leading order, eq.~\eqref{uniCon} evaluates to
\begin{equation}
\begin{split}
A^3\,\Psi_n(u)\Psi'_n(u) \;\sim\; &\left[a_\beta\cos\left(\beta H_0u\right) + b_\beta\sin\left(\beta H_0u\right)\right]\times \\
&\left[\left(3a_\beta + 2b_\beta\beta\right)\cos\left(\beta H_0u\right) + \left(3b_\beta - 2a_\beta\beta\right)\sin\left(\beta H_0u\right)\right]\;. 
\end{split}
\end{equation}
There is no non-trivial combination of $a_\beta$, $b_\beta$, and $\beta\in\Reals^+$ that prevents these plane waves from oscillating out of control in the limit $u\rightarrow\infty$, meaning the only option to satisfy \eqref{uniCon} is setting $a_\beta=0=b_\beta$. This eliminates the massive continuum states from the spectrum. In fact, one can arrive at the same results by requiring the \textit{normalizability} of the wavefunctions, which likewise entirely excludes the continuum modes.
For illustration, we show the correctly normalized wavefunctions in Fig.~\ref{fig:yplot}.

In summary, we find that that the general solution \eqref{gensolu} paired with the unitary boundary condition \eqref{uniCon} implies that
the decomposition of a general, physical graviton mode (\ref{KKdecomp}) is given by
\begin{align}\label{KKfixed}
\tilde{h}_{\mu\nu}(x^\mu,u) = a_0\,\hKK{0}_{\mu\nu}(x^\mu) + a_1\cosh(H_0u - c_u)\,\hKK{1}_{\mu\nu}(x^\mu) \,.
\end{align}
$\hKK{0}_{\mu\nu}(x^\mu)$ here corresponds to a 4D graviton mode with mass parameter $m_0=0$, while $\hKK{1}_{\mu\nu}(x^\mu)$
corresponds to a 4D graviton mode separated by a mass gap $m_1=\sqrt{2}H_0$. Both modes are also subject to an offset in mass, $\delta m=\sqrt{2}H_0$,
which corresponds to the usual background contribution in de Sitter space-time. This kind of graviton spectrum might have been anticipated, given that it is qualitatively similar to 
the previously cited works \cite{Herrera-Aguilar:2009azq,Barbosa-Cendejas:2007ucz}, whose setups also feature a (related but different) modified P\"oschl-Teller potential. 
The key difference here is that the continuous spectrum of graviton modes (separated by a mass gap $m_\beta\geq3/2H_0$) is inconsistent 
with the boundary conditions at the horizon and is therefore unphysical. This can also be seen from the fact that the 5D wavefunctions of the continuum graviton modes 
are not normalizable. We will in the following set $a_0=1$ and $a_1=1$ (noting that $\cosh(-c_u)=1+H_0^2/(2\mu^2)+\mathcal{O}(H_0/\mu)^4\approx1$)
in order to obtain canonically normalized graviton states on the visible brane.\footnote{%
\label{foot:normalization}%
Normalizing the extra dimensional wave functions in this way de facto seems to be standard in the literature. 
However, it should be noted that the solutions of~\eqref{Schrod}, $\psi_n(u)$ in~\eqref{gensolu}, are orthogonal only when integrated over a complete $u$ domain.
Once interpreted as 5D wavefunctions of the graviton states on $z$, we posit that the strictly more correct procedure is to normalize the 
graviton modes to their respectively causally connected domains in the whole extra dimension, though picking a physically correct normalization may depend on the history of the horizon formation. We perform such a normalization in detail in App.~\ref{sec:normalization} and find that the resulting modifications lead to mass-mixing of the graviton
states and modifications of their masses, however, these modifications are completely negligible as they only constitute a difference at the relative order $\mathcal{O}(10^{-31})$.
\nopagebreak}

\subsection{Modified Newtonian potential}
The principal experimental signature of our model is the modification of 4D gravitational phenomenology, namely, changes to the 4D Newtonian potential. 
Modifications to the Newtonian potential from KK graviton states in models with a static 4D cosmology are well-known even in terms of general $\Psi_n$ \cite{Brandhuber:1999hb,Arkani-Hamed:1999wga,Csaki:2000fc}, yielding
\begin{align}
U(r) \sim -\sum_n|\Psi_n(0)|^2\frac{e^{-m_nr}}{r} \,,
\end{align}
where $r$ is the radial coordinate in the three dimensional space of the visible brane. This assumes that the respective wavefunctions $\Psi_n(u)$ are normalized and orthogonal, which is 
an excellent approximation here (see the discussion in footnote~\ref{foot:normalization} and App.~\ref{sec:normalization}). For completeness, we repeat the derivation here for the present case of an FLRW metric on the branes.

The calculation can be done as in standard weak-field GR, where the scalar potential may be identified with the (00) component of the graviton \cite{Carroll2014}.
Hence, we begin with the graviton equation of motion \eqref{EOMlin}, write $\tilde{h}_{00}(x^a) = U(x^a)$, and include a source term corresponding to a point mass $M_s$ on the visible brane at $r=0$ \cite{Dai:2012ni},
\begin{align} \label{EOMU}
\left(\tilde{\nabla}_a\tilde{\nabla}^a + \frac{3\tilde{\nabla}_aA\tilde{\nabla}^a}{A} + 2H_0^2\right)U = \frac{M_s\delta(r)\delta(u)}{M_p^2\sqrt{\tilde{g}}}\;.
\end{align}
We are interested in static and spherically symmetric solutions $(\partial_tU=\partial_\theta U=\partial_\phi U=0)$, and similarly to the KK decomposition employed in the previous section, we may expand $U$ in terms of previous calculated wavefunctions as
\begin{align}
U(r,u) = \sum_{n=0,1}U_n(r)\Psi_n(u) \,.
\end{align}
After converting $x^i$ to spherical coordinates, this leaves us with
\begin{align}
\Psi_n\left(\!-\,\partial_r^2 - \frac{2\partial_r}{r} + 2H_0^2\right)U_n + U_n\left(\!-\,\partial_u^2 - \frac{3\partial_uA\partial_u}{A}\right)\Psi_n = \frac{M_s\delta(r)\delta(u)}{M_p^2} \,.
\end{align}
By construction, the $\Psi_n$ are mass eigenfunctions of the operator $-\partial_u^2 - 3(\partial_uA/A)\partial_u$.
Integrating over $u$ while using the orthogonality of the eigenfunctions and $\Psi_n(0)\approx1$ by the previously established normalization, 
the remaining differential equation in $r$ can be solved (using the boundary condition $U_n(\infty)=0$) to obtain the result
\begin{align}
U(r,0) = -\frac{M_s}{M_p^2}\sum_{n=0,1}\frac{e^{-\sqrt{2H_0^2 + m_n^2}r}}{r} \,.
\end{align}
The first ($m_0=0$) term in this sum matches the Newtonian potential predicted by GR in dS space. The second term ($m_1=\sqrt{2} H_0$) corrects this result and represents new 4D physics that result from the 5D geometry.
For short distances, this implies a standard $r^{-1}$ Newtonian potential enhanced by a factor of two, though such a modification can easily be accommodated for by rescaling the the numerical value of the Planck mass by a corresponding factor $M_p\rightarrow \tilde{M}_p=\sqrt{2}M_p\approx3.4\times10^{18}\,\mathrm{GeV}$. 
We have not mentioned this rescaling in Sec.~\ref{sec:HPandCC} so as to not overload the discussion. In hindsight, we may have adjusted $R$ so as to yield 
$\tilde{M}_p$ in \eqref{eq:Mpl}, corresponding to a numerical value $\mu\pi R\approx243.56$ (as compared to $243.21$ for the standard Planck mass),
though this represents a negligible change to the discussion as a whole.

At astronomical distances $r\gtrsim1/H_0$ the Newtonian potential is damped and the extra massive mode amplifies this effect.  Nonetheless, the mass of the second mode is so small that it avoids experimental bounds on massive spin-2 fields on the low side. Current bounds on massive gravitons are collected in~\cite{ParticleDataGroup:2022pth} and reviewed together with future projections in~\cite{deRham:2016nuf}. The most reliable and model-independent limit on the graviton mass apparently comes from LIGO/Virgo and relies only on a modified graviton dispersion relation to constrain the kinematic graviton mass to $m_h\leq1.27\times10^{-23}\,\mathrm{eV}$ \cite{LIGOScientific:2021sio}. There are more stringent bounds that constrain the Yukawa dampening of the gravitational potential on scales comparable to the solar system \cite{Will:2018gku} or galaxy clusters \cite{Gupta:2018hgm, Piorkowska-Kurpas:2022agy}. Those put bounds as low as $m_h\lesssim5.63\times10^{-24}\,\mathrm{eV}$ or $m_h\lesssim4.7\times10^{-30}\,\mathrm{eV}$, respectively.  The most stringent constraints seem to arise from weak lensing~\cite{Choudhury:2002pu} and effects of the graviton mass on the cosmological evolution~\cite{Jusufi:2023xoa,Gonzalez:2023rsd} which reach deep into the $m_h\lesssim10^{-32}\,\mathrm{eV}$ region but become increasingly model dependent.\footnote{%
In fact, the recent~\cite{Gonzalez:2023rsd} claims a detection with $m_h=2.4_{-0.7}^{+2.1}\times10^{-33}\,\mathrm{eV}$ which would indeed fit our mode $m_1=2.1\times10^{-33}\,\mathrm{eV}$. However, the scenario of~\cite{Gonzalez:2023rsd} is based on non-standard cosmology~\cite{Jusufi:2023xoa} and has additional free parameters. It remains to be seen whether such a measurement can be upheld by alternative means.}
Another idea is that the presence of extra dimensions itself may lead to an alteration of the propagation of gravitational waves, see e.g.~\cite{Andriot:2019hay,Andriot:2021gwv}, and additional contributions to the gravitational memory effect from new light degrees of freedom~\cite{Ferko:2021bym} but the details would have to be worked out specifically for our model.

In the future, space-based gravitational wave interferometers such as LISA~\cite{LISA:2017,LISA:2022jok} as well as the ground-based Einstein telescope~\cite{Maggiore:2019uih,ET:2020} will sharpen the direct and indirect graviton mass constraints with enormously improved statistics of multi-messenger binary mergers
and more refined constraints on the stochastic gravitational wave background. Likewise, improved astronomical observations of galaxy clusters and potentially superclusters, as well as improved cosmological constraints leave us optimistic that our scenario can ultimately be tested in the foreseeable future.

Finally, we mention that there are also theoretical bounds to consider, namely, the Higuchi bound~\cite{Higuchi:1986py}.\footnote{%
Application of this bound to the present scenario is a bit subtle, since it relies on the standard Friedmann equations to link the Hubble constant to a positive de Sitter curvature, whereas in the present setup, there is no explicit constant energy density in 4D beyond the brane tension (in the case $\rho_0=0$), with the Hubble rate $H_0$ being fundamentally linked instead to the negative 5D cosmological constant.} This constrains $m_h^2 \geq 2H_0^2$ for massive gravity in 4D de Sitter space in order to avoid the appearance of negative norm states after quantization. Our lightest graviton state saturates this bound. Going beyond the present approximation of a cosmological constant dominated equation of state may slightly change the graviton masses.

\section{Time-dependent energy densities and brane cosmology} \label{sec:cosmo}
The discussions of Sec.~\ref{sec:NKKK} and Sec.~\ref{sec:KKstates} correspond to branes dominated 
by constant energy densities and thus imply a simple vacuum energy dominated 4D cosmology. 
In this final section we show how it is possible to generalize our results to also account for the presence of perfect fluids with time-dependent energy densities $\rho_{0,\pi}(t)$ and pressure $p_{0,\pi}(t)$ on the branes,
as already included in our general treatment of Sec.~\ref{sec:general}. This allows for 4D cosmologies that appear radiation or matter dominated, 
despite the fact that \mbox{$\rho_{0,\pi}(t),p_{0,\pi}(t)\ll\lambda_{0,\pi}$} still represent only extremely small corrections to the four-dimensional brane tensions.
It is in fact because of this feature that we can work in perturbation theory around the vacuum energy dominated case in the 5D picture, despite the fact that the 4D picture changes from vacuum energy, to radiation or matter domination.

We start with the general metric \eqref{genmet} and solution \eqref{eq:JKAnsatz}, and assume a static extra dimension $\dot{\Bf}=0$.\footnote{%
While it should also be possible to find a more general adiabatic solution ($\dot{\Bf}/\Bf\ll\dot{\Af}/\Af$) for time-dependent matter distributions,
we expect quantitative differences to be small as compared to our treatment, simply because we stay perturbatively close to the exact solutions
of the previous section.}
This leaves us with the metric
\begin{align} \label{genmet2}
\mathrm{d}s^2 = \Nf^2(t,z)\dd t^2 - \Af^2(t,z)\delta_{ij}\dd x^i \dd x^j - \dd z^2 \,.
\end{align}
Making no a-priori assumptions about the separability of $\Nf(t,z)$ and $\Af(t,z)$, one finds that the $(00)$, $(ii)$, $(55)$, and $(05)$ bulk Einstein equations 
are all non-degenerate. To find a solution for $\Nf$ and $\Af$ that produces viable Friedmann equations, we employ an ansatz similar to that used in \cite{Csaki:1999mp} where we expand around a ``base-case'' time-independent solution, which for us is the NKKK warp factor \eqref{NKKKmet}. 
In terms of the general perturbative functions $\delta a_i(t)$, $\delta N_i(z)$, and $\delta A_i(z)$, our ansatz reads
\begin{align}\label{eq:ansatz}
&\Nf(t,z) = A(z)\left[1 + \sum_{i=1}^\infty \delta a_i(t) \delta N_i(z)\right],& &\Af(t,z) = a(t)A(z)\left[1 + \sum_{i=1}^\infty \delta a_i(t) \delta A_i(z)\right].&
\end{align}
In \cite{Csaki:1999mp} the base-case was taken to be the well-known RS1 warp factor $A(z)=\exp(-\mu z)$, but the NKKK metric is naturally the more suitable starting point here. After inserting the ansatz above into the jump conditions \eqref{eq:0bc},
one finds that $\delta a_1(t) \propto \rho_0(t)$, and since we have that $\rho_0(t) \ll \lambda_0$, it suffices to work to leading order 
in $\rho_0(t)$, thus implying that we can neglect higher order terms in \eqref{eq:ansatz}. The ansatz \eqref{genmet2} then simplifies to
\begin{align} \label{eq:Nans}
\Nf(t,z) &= \frac{\sinh(c_z - \mu z)}{\sinh(c_z)}\left[1 + \frac{\rho_0(t)}{\lambda_0}\delta N(z)\right],&  \\ \label{eq:Aans}
\Af(t,z) &= a(t)\frac{\sinh(c_z - \mu z)}{\sinh(c_z)}\left[1 + \frac{\rho_0(t)}{\lambda_0}\delta A(z)\right] \,.& 
\end{align} 
Inserting \eqref{eq:Nans} and \eqref{eq:Aans} into \eqref{eq:0bc}, one finds that the two functions must respect the boundary conditions
\begin{align}
\delta N'(0) &= \frac{\lambda_0}{6M^3}\left(2 + 3w\right)\;,& \delta A'(0) &= -\frac{\lambda_0}{6M^3}\;,&
\end{align}
where we have also required that $\delta N(0)=\delta A(0)=0$ in order to ensure a smooth connection with the base-case solution.
Assuming that $\delta N(z)$ and $\delta A(z)$ have a similar $z$-dependence, we can infer from the above conditions that
\begin{equation}
\delta N(z)=-(2 + 3w)\delta A(z)\;.
\end{equation}
This leaves us with a single unknown function to be determined from the Einstein equations.

Taking
\begin{align}
\delta A(z) = -\frac{\lambda_0}{6\mu M^3}\frac{\sinh(\mu z)\sinh(c_z)}{\sinh(c_z - \mu z)} \,,
\end{align}
the $(00)$ and $(ii)$ equations yield relations between the Hubble function and energy density that are reminiscent of the 4D Friedmann equations to first order in $\rho_0$, but with an additional (typically highly suppressed) $\rho_0^2$ contribution,\footnote{%
One can also obtain eq.~\eqref{eFE1} by formally replacing $\lambda_0\rightarrow\lambda_0+\rho_0(t)$ in eq.~\eqref{HDefs}. Since all of these equations originate from the general boundary conditions~\eqref{eq:0bc}, which do not require taking time derivatives, we expect these relations to hold for even the most general metrics.}
\begin{align}\label{eFE1}
\mathcal{H}^2_0(t) &= H_0^2 + \frac{\sqrt{H_0^2 + \mu^2}}{3M^3} \rho_0(t) + \frac{\rho^2_0(t)}{36M^6}\;,& \\\label{eFE2}
\dot{\mathcal{H}}_0(t) &= -(1 + w)\left(\frac{\sqrt{H_0^2 + \mu^2}}{2M^3} \rho_0(t) + \frac{\rho^2_0(t)}{12M^6}\right)\;.
\end{align}
We also encounter the familiar energy conservation relation
\begin{align}
\dot{\rho}_0(t) = -3(1 + w)\mathcal{H}_0(t)\rho_0(t) \,,
\end{align}
which follows from the general conservation equation \eqref{econGen} and the $(05)$ Einstein equation.

A clear connection to the standard 4D Friedmann equations can be made by defining the rescaled energy densities
\begin{align}
\hat{\rho}_{0,\pi}(t) := \frac{\mu M_p^2}{M^3}\rho_{0,\pi}(t) \,,
\end{align}
with an (here assumed to be time independent) effective 4D Planck mass $M_p$.
With this, the Friedmann equations \eqref{eFE1} and \eqref{eFE2} are given by
\begin{align}\label{eq:eFE3}
\mathcal{H}^2_0(t) &= H_0^2 + \frac{\hat{\rho}_0(t)}{3M_p^2} + \frac{\hat{\rho}^2_0(t)}{36\mu^2M_p^4} \,,& 
\dot{\mathcal{H}}_0(t) &= -(1 + w)\left(\frac{\hat{\rho}_0(t)}{2M_p^2} + \frac{\hat{\rho}^2_0(t)}{12\mu^2M_p^4}\right) \,.&
\end{align}
The quadratic corrections to the standard Friedmann equations are unimportant as long as $\rho_{0,\pi}\ll\lambda_0$  and play a role only if $\rho_{0,\pi}\sim \lambda_0\sim M^3\mu$. Crucially, these equations reduce to $\mathcal{H}_0(t)=H_0$ and $\dot{\mathcal{H}}_0(t)=0$ in the cosmological constant dominated regime, as one should expect. 

It should also be noted that, even though we have derived the Friedmann equations in terms of the visible brane quantities $\rho_0$ and $p_0=w\rho_0$, 
one may also use the hidden brane jump conditions as a starting point to obtain the analogous hidden brane Friedmann equations. 
Furthermore, if we assume that the constant brane tensions $\lambda_{0,\pi}$ are fixed relative to one another as in the base case \eqref{ABCs}, 
we may evaluate the general jump conditions with the present metric to find a relation between the time-dependent energy densities on each brane,
\begin{align}\label{eq:nonlocalFT}
\rho_0(t) = A^2(\pi R)\rho_\pi(t) + \Ord\big(\rho_\pi^2\big) \,,
\end{align} 
which may in turn be used to show agreement between the visible and hidden brane Friedmann equations.
Just as in the tightly fixed relation between $\lambda_0$ and $\lambda_\pi$ that is required in order to obtain a correct $M_p$ in the $\rho_{0,\pi}=0$ case, 
one may interpret eq.~\eqref{eq:nonlocalFT} as a non-local fine-tuning of energy densities on the branes originating from the requirement $\dot{R}=0$. If this requirement is dropped, 
the more general solution will have the feature that whatever the energy density and equation of state on the branes may be, the radii $R_0$ and $R_\pi$, and hence $R$, 
would (adiabatically) adjust themselves in such a way as to satisfy the equations of motions, thereby also adjusting themselves to fulfill the non-local ``tuning''.

Finally, we discuss a problem in turning this model into a fully realistic cosmology. While eqs.~\eqref{eFE1} and~\eqref{eFE2} take the form of the standard Friedmann equations for \mbox{$\rho_0(t)\ll \lambda_0 \sim M^3\mu$}, we note that the expansion rate is not suppressed by the 4D Planck mass $M_p$ but only by the 5D scale $M$ (alternatively, if $M_p$ is introduced such as in  eqs.~\eqref{eq:eFE3}, it is not the actual 4D brane energy density that is appearing but the upscaled $\hat\rho_0$).
The fact that the ``Planck mass'' appearing in the expansion law differs from the actual Planck mass obtained by integrating out the extra dimension (which affects the strength of gravity on the branes and tensor mode perturbations) is nothing unusual. For example, this is at the heart of the suppression or enhancement of the tensor-to-scalar ratio of gravitational perturbations in scenarios of braneworld inflation~\cite{Langlois:2000ns,Giudice:2002vh,Frolov:2002qm} (see~\cite[ch.\ 5.1]{Maartens:2010ar} for a review).
We are however in a more extreme situation here with the splitting between $M$ and $M_p$ being as large as it needs to be in order to explain the hierarchy between electroweak and Planck scale. One might of course simply tune $M^3/\mu$ to $M_p^2$, however this inherently restores the original hierarchy problem as a tuning between the bulk cosmological constant $\Lambda$ and $M$, so we discard that option. Hence, though our derived expansion law is qualitatively identical to the standard Friedmann law, it delivers a quantitatively different connection between $\mathcal{H}_0(t)$ and $\rho_0(t)$. This leads to a problem because both the \textit{absolute} value of the expansion rate $H_0$ and the \textit{absolute} value of the dynamical brane energy density $\rho_0(t)$ (not $\hat{\rho}_0(t)$) can independently be inferred by observations associated to different cosmological epochs.

In contrast to the observable $\mathcal{H}_0(t)$ and $\rho_0(t)$, we note that the brane energy density $\lambda_0$ is not an observable and may only be inferred based on the corresponding (observable) Hubble rate. Hence, the problem described above could be resolved if there were a mechanism that dynamically tunes $\lambda_0$ in order to obtain a correct relation between $\mathcal{H}_0(t_0)$ and $\rho_0(t_0)$, however, such an adjustment must be applied for each cosmological epoch $t_0$. To make this statement explicit, we can write eq.~\eqref{eFE1} as
\begin{equation}\label{eq:tuning}
 \mathcal{H}_0^2(t)~=~\frac{1}{36\,M^6}\left[\lambda_0^2+2\,\lambda_0\rho_0(t)+\rho_0^2(t)-36\mu^2M^6\right]\;.
\end{equation}
Given a requirement on the absolute sizes of $\mathcal{H}_0(t_0)$ and $\rho_0(t_0)$ at a certain epoch $t_0$, $\lambda_0$ can obviously be tuned so as to fulfill this equation and accommodate a realistic cosmological expansion during that epoch. 
The exact solution to \eqref{eq:tuning} of this kind is given by
\begin{equation}
\lambda_0=6M^3\mu\sqrt{1+\frac{\mathcal{H}^2(t_0)}{\mu^2}}-\rho_0(t_0) \,,
\end{equation}
and one clearly sees that, for all interesting epochs, the tuning corresponds to a only minuscule deviation from the approximate value $\lambda_0\approx6M^3\mu$. Interestingly, $\lambda_0>6M^3\mu$
for cosmological constant dominated epochs, while $\lambda_0<6M^3\mu$ if there are additional dynamical brane fluids.
Unfortunately, as soon as we let the solution evolve for $t>t_0$ and $\rho(t)<\rho(t_0)$, not only will the expansion rate quickly start to differ from potentially realistic values as we go away from $t_0$, but we are eventually also driven into a regime where $\mathcal{H}_0^2$ would have to become negative, rendering our solution unphysical --
we recall that our solution in general only makes sense if $\sigma_0>1$, implying here strict positivity of the square bracket in eq.~\eqref{eq:tuning}.\footnote{%
We remark that a negative sign in $\mathcal{H}_0^2$ can be compensated by allowing $\mu$ to turn imaginary. This transformation relates our solution for $\Lambda<0$ to the solution with $\Lambda>0$ (see e.g.\ eq.~(18) of~\cite{Im:2017eju} and the discussion in~\cite{Kaloper:1999sm}). This suggests that the two solutions could be dynamically connected depending on the momentary value of the evolving brane energy densities $\rho_{0,\pi}(t)$. This connection is interesting as it may allow one to understand how solutions with discretely different $\lambda_0$ are connected, though a detailed investigation of this possibility is beyond the scope of this work.} 

The only way out of this dilemma is to deviate from our quasi static approximation and allow $\lambda_0$ to evolve extremely slowly, implying that the size of the extra dimension should also vary. However, the physical mechanism behind such a dynamical evolution of $\lambda_0$ is presently unclear (in principle the evolution of $\lambda_0$ does not have to be continuous but could also happen step-wise), such that in the given form our model can only realistically be applied to the current epoch of cosmological constant domination. We note that in order for any dynamical tuning mechanism to work, $\lambda_0$ must start with a value \textit{smaller} than $6M^3\mu$ and evolves towards this most natural value in the course of cosmic evolution.
That is, the value of $\delta\lambda_0/\lambda_0\approx10^{-88}$ required for a realistic expansion rate today might be obtained dynamically in a natural way as it is close to a fixed point of the extra dimensional evolution. Interestingly, having the brane tensions evolve in such a way may also allow for an explanation to the cosmological ``why now'' problem: 
The perceived 4D cosmological constant is continuously evolving but can only become the dominant source of expansion once $\lambda_0$ crosses from $\lambda_0<6M^3\mu$ to $\lambda_0>6M^3\mu$,
at which point the dynamical evolution of $\lambda_0$ halts and 4D expansion is quickly dominated by a cosmological constant. 
This crossing effect is somewhat reminiscent of the relaxion solution to the hierarchy problem~\cite{Graham:2015cka} inspired by Abbott's solution to the cosmological constant problem~\cite{Abbott:1984qf}. 
However, a solution with evolving ``constant'' brane tensions would require an ansatz with $\dot{\mathcal{B}}\neq0$, i.e.\ deviating from the quasi-static approximation we have taken from the beginning
to expose the potential, excitations, and eventually possible tunnelings of the radion mode. While this is beyond the scope of this work we think it is an interesting problem to return to in the future.

\section{Conclusions}\enlargethispage{0.5cm}
We have discussed an extra dimensional braneworld scenario with hyperbolic warping and a bulk horizon.
For strictly constant brane energy densities, the corresponding metric was discussed earlier~\cite{Nihei:1999mt,Kaloper:1999sm,Kim:1999ja}. We have for the first time taken seriously the possibility of extending the spacetime beyond the bulk horizon and have shown that this allows us to relate the splitting between the 4D Planck mass and 4D cosmological constant to an order one ratio \mbox{$\gamma=R/R_0\approx2.34$} of proper distances in the extra dimension.
The required fine tuning is not reduced but promoted to a fine tuning of brane tensions that must be extremely close to their most natural values (corresponding to the electroweak scale for a solution to the hierarchy problem).
In order to make sure that this scenario obeys causality and unitarity, we have analytically computed the Kaluza-Klein spectrum of 4D graviton modes and their boundary conditions
at the horizon. This lead us to a modified P\"oschl-Teller potential whose solutions predict the expected massless graviton as well as an additional massive, but light graviton mode with $m_1=\sqrt{2}H_0$, which is a smoking gun signature of this scenario that could be tested for in the future. The close proximity of this mode to the usual massless graviton may also give rise to phenomena like graviton mixing and graviton oscillations.

An important merit of our solution is that all involved brane energy densities are positive (unlike, for example, in the usual RS1 model). We have also studied the dynamical evolution of brane energy densities modeled by perfect fluids on the branes in this geometry.
We have found that a conventional 4D Friedmann expansion law can be obtained, but realistic combinations of expansion rate and dynamically evolving 4D energy densities (as measured in the real world) typically only exist in a fine tuned region of parameters (corresponding to the fine tuning of constant 4D brane energy densities in the cosmological constant dominated case). Requiring a healthy cosmological evolution at all epochs points us towards slowly running brane tensions -- an extension of our solution that must be further clarified in the future. The adiabatically slow evolution of the size of the 5th dimension and associated slow running of the 4D brane tensions should also be further studied by including perturbations to the (runaway) radion mode, which we have shown to be non-dynamical in the adiabatic case. Interesting variations of this model include the extension to more than one extra dimension, multiple causally disconnected regions, and/or the inclusion of fermionic bulk modes.

\section*{Acknowledgments}
We would like to thank Tony Gherghetta for useful discussions and Andreas Bally for a careful reading of the manuscript.
We have made excessive use of the \href{http://www.xact.es}{\texttt{xAct}} package for Wolfram \texttt{Mathematica} to perform calculations in this work, in particular the packages~\cite{Martingarcia:2008,Brizuela:2008ra,Nutma:2013zea,Frob:2020gdh}.
The work of AT was in part performed at the Aspen Center for Physics, which is supported by National Science Foundation grant PHY-1607611.
The work of AT in Aspen was partially supported by a grant from the Simons Foundation.

\appendix
\section*{Appendix}
\section{Hierarchy Problem} \label{sec:HP}
To render our paper self-contained, we briefly recall the details of the how the hierarchy problem is solved in a scenario like ours. 
As in the RS1 scenario, the 4D Planck mass $M_p$ is not a fundamental scale, but rather an effective scale, derived from the fundamental 5D mass scale $M$,
and potentially subject to a large enhancement due to warping. 

Including the Higgs field $\Phi(x^\mu)$ that depends only on the 4D coordinates $x^\mu$ into $\Lag_0$ and $\Lag_\pi$, 
the relevant part of the full 5D action \eqref{action} reads  
\begin{align} \label{EHiggsAction}
S \supset \int\dd ^4x\dd z\sqrt{g}\left[\frac{M^3}{2}\mathcal{R} + \big(\delta(z) + \delta(z - \pi R)\big)\!\left(\frac12 g^{\mu\nu}D_\mu\Phi^\dagger D_\nu\Phi - \lambda \big(|\Phi|^2 - v_{\mathrm{EW}}^2\big)^2\right)\right]\,.
\end{align}
Using \eqref{eq:dimred} to derive the effective 4D action on the branes results in 
\begin{align}
S \supset \int\dd ^4x \left(\frac{A^2(z)}{2}D_\mu\Phi^\dagger D^\mu\Phi - A^4(z)\lambda\big(|\Phi|^2 - v_{\mathrm{EW}}^2\big)^2\right)\Bigg|_{z=0,\pi R}\,,
\end{align}
where one needs to insert either $0$ or $\pi R$ for $z$ depending on which brane is being considered. 
Canonically normalizing $\Phi$ as $\hat{\Phi}=A(z)\Phi$ implies that the fundamental physically measurable electroweak scale on the brane 
is given by $\hat{v}_{\mathrm{EW}}=A(z)v_{\mathrm{EW}}$. Thus, since $A(0)=1$, an observer confined to the visible brane determines $\hat{v}_{\mathrm{EW}}\sim 250\,\mathrm{GeV}$
to be the fundamental physical scale, while an observer on the hidden brane is subject to the scaled value $\hat{v}_{\mathrm{EW}}=A(\pi R)v_{\mathrm{EW}}$. 
This is precisely the same mechanism as in the original Randall-Sundrum model \cite{Randall:1999ee}, with the small caveat that RS assumes that our physical world lies on the UV brane away from the origin. We find it more convenient in our model to have the visible brane located at $z=0$, where the experimentally determined Higgs VEV corresponds to the true fundamental scale, a choice that does not alter any of the physics. 

\section{Normalization of the 4D graviton states} \label{sec:normalization}
Here we perform what we deem to be the strictly correct normalization of the extra-dimensional wavefunctions $\Psi_{n=0,1}(u)$ of Sec.~\ref{sec:KKstates}, see Eqs.~\eqref{eq:Psi0} and \eqref{eq:Psi1}. 
Expanding the bulk action (\ref{SConf}) to second order in $\tilde{h}_{ab}$ we arrive at
\begin{align} \label{S5lin}
S_\text{bulk}^\text{(lin)} = M^3\int\dd^4x\dd u\sqrt{\tilde{g}}A^3\left(\tilde{h}_{ab}\tilde{\nabla}_c\tilde{\nabla}^c\tilde{h}^{ab} + \frac{3}{A}\tilde{\nabla}_cA\,\tilde{h}_{ab}\tilde{\nabla}^c\tilde{h}^{ab} + 2H_0^2\tilde{h}_{ab}\tilde{h}^{ab}\right) \,,
\end{align}
where we have employed the transverse-traceless gauge (\ref{ttgauge}). We note that the EOM for $\tilde{h}_{ab}$ obtained from this action precisely matches (\ref{EOMlin}) and we recall that the $H_0^2$ term is not an actual mass but rather appears as a result of expanding around a dS background, commuting covariant derivatives, and using relations like $\tilde{h}^{ab}\tilde{h}^{cd}\mathcal{R}_{acbd}=-H_0^2\tilde{h}_{ab}\tilde{h}^{ab}$. Indeed, the action above describes a massless graviton in dS space as expected.

In order to integrate over the extra dimension, we separate all of the $u$-dependence in the action above by expanding $\tilde{h}_{ab}$ in terms of the KK decomposition (\ref{KKfixed}) and using that $\tilde{h}_{a5}=0$ to write $\sqrt{\tilde{g}}\tilde{\nabla}_c\hKK{n}_{ab}\tilde{\nabla}^c\hKK{n}^{ab}=\sqrt{-\hat{g}}\hat{\nabla}_\rho\hKK{n}_{\mu\nu}\hat{\nabla}^\rho\hKK{n}^{\mu\nu}$, $\sqrt{\tilde{g}}\hKK{n}_{ab}\hKK{n}^{ab}=\sqrt{-\hat{g}}\hKK{n}_{\mu\nu}\hKK{n}^{\mu\nu}$, etc., where hats indicate 4D operators that live on the branes with Greek indices raised and lowered by the effective 4D FLRW metric $\hat{g}_{\mu\nu}$. We make the crucial assumption that the massless graviton mode corresponding to $\Psi_0$ represents the global effect of gravity (picking up effects from across the horizon), while the excited and potentially propagating graviton mode corresponding to $\Psi_1$ must be confined to the respective causally connected regions separated by the horizon. Practically, this is implemented by integrating over the entire extra dimension after changing back to the physical $z$ coordinates with the definition (\ref{eq:uz}) and taking \begin{align}
&\Psi_0\big(u(z)\big) = a_0& \Psi_1\big(u(z)\big) = a_1\coth(\mu z - c_z)\theta\big(c_z/\mu - z\big) \,.
\end{align}
Here, $\theta(c_z/\mu - z)$ is the Heaviside theta function that enforces the support of $\Psi_1$ in the causally connected region. 
We may then perform the integral over $z$ to find the effective 4D action
\begin{align} \label{S4D}
S_\text{brane}^{\mathrm{(lin)}} = M_p^2\int\dd^4x\sqrt{-\hat{g}}\bigg[&a_0^2k_{00}\Big(\hat{\nabla}_\rho\hKK{0}_{\mu\nu}\hat{\nabla}^\rho\hKK{0}^{\mu\nu} - 2H_0^2\hKK{0}_{\mu\nu}\hKK{0}^{\mu\nu}\Big) \nonumber \\
&+ 2a_0a_1k_{01}\Big(\hat{\nabla}_\rho\hKK{0}_{\mu\nu}\hat{\nabla}^\rho\hKK{1}^{\mu\nu} - 3H_0^2\hKK{0}_{\mu\nu}\hKK{1}^{\mu\nu}\Big) \nonumber \\
&+ a_1^2k_{11}\Big(\hat{\nabla}_\rho\hKK{1}_{\mu\nu}\hat{\nabla}^\rho\hKK{1}^{\mu\nu} - 4H_0^2\hKK{1}_{\mu\nu}\hKK{1}^{\mu\nu}\Big)\bigg] \,,
\end{align}
where the $k_{ij}$ represent the results of the integrations and are given by
\begin{align} \label{kijDefs}
&k_{00} = 1\;,& &k_{01} = -\frac{M^3}{\mu M_p^2}\;,& &k_{11} = \frac{H_0M^3}{\mu^3M_p^2}\big(\mu\cosh(c_z) + H_0\,c_z\big)\;.
\end{align}
We recall that $M_p$ and $M$ are related through the relation (\ref{MpM}).

The presence of the off-diagonal $\hKK{0}$--$\hKK{1}$ terms in this action has potentially important consequences that must be addressed. One may naively assume that these terms cancel since the Schrödinger eigenstates $\psi_n$ are orthogonal, however, this is not necessarily the case for the full $u$-dependent part of the 5D graviton, $\Psi_n=A^{-3/2}\psi_n$, that enters into the extra dimensional integral. We must thus diagonalize the 4D action (\ref{S4D}) to recover the true masses of each mode and their mixing parameter. We note that if the mixing term would (artificially) be dropped, the predicted mass values $m_0^2=0$ and $m_1^2=2H_0^2$ are reflected in the action above which receives the same kind dS background contributions as the 5D graviton action i.e.\ the resulting EOMs would take the form
\begin{align} \label{dSKGeq}
\Big(\hat{\nabla}_\rho\hat{\nabla}^\rho + 2H_0^2 + m_n^2\Big)\hKK{n}_{\mu\nu} = 0 \,.
\end{align}

We may simultaneously normalize and diagonalize the 4D action (\ref{S4D}) with the field redefinitions
\begin{align} \label{hredefs}
&\hKK{0}_{\mu\nu} = \frac{\sqrt{1 + \varepsilon}}{\sqrt{2}\,a_0}\Big(\sqrt{2 + \varepsilon}\,\hKKp{0}_{\mu\nu} + \sqrt{\varepsilon}\,\hKKp{1}_{\mu\nu}\Big)\;,& &\hKK{1}_{\mu\nu} = \frac{\sqrt{1 + \varepsilon}}{\sqrt{2k_{11}}\,a_0}\Big(\sqrt{\varepsilon}\,\hKKp{0}_{\mu\nu} + \sqrt{2 + \varepsilon}\,\hKKp{1}_{\mu\nu}\Big) \,,
\end{align}
where for convenience we have defined the parameter
\begin{align} \label{epDef}
\varepsilon = \frac{1}{\sqrt{1 - k_{01}^2/k_{11}}} - 1 \,.
\end{align}
Inserting the definitions above into (\ref{S4D}) yields the diagonalized 4D action
\begin{align} \label{S4Dp}
S_\text{brane}^{\mathrm{(lin)}} = M_p^2\int\dd^4x\sqrt{-\hat{g}}\Big(&\hat{\nabla}_\rho\hKKp{0}_{\mu\nu}\hat{\nabla}^\rho\hKKp{0}^{\mu\nu} - (2 - \varepsilon)H_0^2\hKKp{0}_{\mu\nu}\hKKp{0}^{\mu\nu} \nonumber \\
&+ \hat{\nabla}_\rho\hKKp{1}_{\mu\nu}\hat{\nabla}^\rho\hKKp{1}^{\mu\nu} - (4 + \varepsilon)H_0^2\hKKp{1}_{\mu\nu}\hKKp{1}^{\mu\nu}\Big)\;.
\end{align}
In the diagonal mass basis the masses for each mode are identified as (again in line with~(\ref{dSKGeq}))
\begin{align}
\frac{{m'_0}^2}{H_0^2} = -\varepsilon \qquad\text{and}\qquad \frac{{m'_1}^2}{H_0^2} = 2 + \varepsilon \,.
\end{align}

The definition for $\varepsilon$ in (\ref{epDef}) is exact and fully determines the physical masses. To get a feel for what values it might realistically take, we may make the realistic assumption that $\mu\gg H_0$ and use the definitions (\ref{kijDefs}) to find that $k_{01}^2/k_{11} \approx M^3/(\mu M_p^2)$. This implies that when $\mu$ and $M$ take hierarchy problem-resolving values of $\approx1\text{ TeV}$, one finds $\varepsilon\approx10^{-31}$ and a nearly perfect match to the originally predicted values of $m_0^2=0$ and $m_1^2=2H_0^2$. It should be noted that since $\varepsilon>0$, the ``massless'' mode is strictly speaking tachyonic, though its mass is \textit{far} below any possibly measurable value and should thus cause no practical problems. For all intents and purposes it indeed represents the expected 4D massless graviton. In our opinion the tachyonic behavior and its relative smallness make sense, 
from the point of view that we are, in principle, discussing the static limit of an adiabatic runaway solution.

\bibliography{library}
\addcontentsline{toc}{section}{Bibliography}
\bibliographystyle{utphys}

\end{document}